\documentclass[twocolumn]{aastex62}
\usepackage{hyperref,graphicx,wrapfig,booktabs,amsmath,comment}

\begin{document}

\title{A Catalog of AGN Host Galaxies Observed with HST/ACS: Correlations between Star Formation and AGN Activity}

\author{Aaron Stemo}
\affiliation{Department of Astrophysical and Planetary Sciences, University of Colorado, Boulder, CO 80309, USA}

\author{Julia M. Comerford}
\affiliation{Department of Astrophysical and Planetary Sciences, University of Colorado, Boulder, CO 80309, USA}

\author{R. Scott Barrows}
\affiliation{Department of Astrophysical and Planetary Sciences, University of Colorado, Boulder, CO 80309, USA}

\author{Daniel Stern}
\affiliation{Jet Propulsion Laboratory, California Institute of Technology, 4800 Oak Grove Drive, Pasadena, CA 91109, USA}

\author{Roberto J. Assef}
\affiliation{N\'{u}cleo de Astronom\'{\i}a de la Facultad de Ingenier\'{\i}a y Ciencias, Universidad Diego Portales, Av. Ej\'{e}rcito Libertador 441, Santiago, Chile}

\author{Roger L. Griffith}
\affiliation{Jet Propulsion Laboratory, California Institute of Technology, 4800 Oak Grove Drive, Pasadena, CA 91109, USA}

\begin{abstract}
We present the Advanced Camera for Surveys Active Galactic Nuclei (ACS-AGN) Catalog, a catalog of 2585 active galactic nucleus (AGN) host galaxies that are at redshifts $0.2<z<2.5$ and that were imaged with the \textit{Hubble Space Telescope's} Advanced Camera for Surveys (ACS). Using the ACS General Catalog (ACS-GC) as our initial sample, we select an AGN subsample using \textit{Spitzer} and \textit{Chandra} data along with their respective established AGN selection criteria. We then gather further multi-wavelength photometric data in order to construct spectral energy distributions (SEDs). Using these SEDs we are able to derive multiple AGN and host galaxy properties, such as star formation rate, AGN luminosity, stellar mass, and nuclear column density. From these data, we show that AGN host galaxies tend to lie below the star-forming main sequence, with X-ray-selected AGN host galaxies being more offset than IR-selected AGN host galaxies. This suggests that there is some process, possibly negative feedback, in AGN host galaxies causing decreased star formation. We also demonstrate that there is a positive trend between star formation rate and AGN luminosity in AGN host galaxies, in individual redshift bins and across all redshift bins, and that both are correlated with the stellar mass of their galaxies. This points towards an underlying link between the stellar mass, stellar growth, and SMBH growth in a galaxy.
\end{abstract}


\section{Introduction} \label{sec: intro}
Supermassive black holes (SMBHs) are thought to exist at the centers of all massive galaxies, and there is evidence of these central SMBHs being coupled to their host galaxies. In particular, there are strong correlations between observed properties of the SMBH and host galaxy properties; for example, the relations between SMBH mass and host galaxy bulge mass ($M_{BH}$ -- $M_{Bulge}$ relation), luminosity ($M_{BH}$ -- $L_{Bulge}$ relation), and velocity dispersion ($M_{BH}$ -- $\sigma_{Bulge}$ relation) \citep[e.g.,][]{Magorrian1998,Ferrarese2000,Gebhardt2000,Marconi2004,Greene2006,Gultekin2009,Kormendy2013}. Further, the cosmic star formation history and black hole accretion history appear to behave similarly, peaking near $z$ $\sim$ 2 and declining to local values \citep[e.g.,][]{Boyle1998,Silverman2008,Shankar2009,Aird2010,Madau2014}. These relations point towards a connection between SMBH growth and host galaxy growth, even though the spatial scales of active SMBH accretion (i.e., an active galactic nucleus, AGN) and host galaxy star formation differ by up to 9 orders of magnitude. 

Despite the large separations in spatial scales, both SMBH growth and star formation require a cold gas supply. This alludes to a connection that may lie in coincident feeding mechanisms. There are two primary modes of star formation: via the star-forming main sequence, or merger driven star formation. Most star-forming galaxies lie along a star-forming main sequence, where the star formation rate is tightly correlated with stellar mass \citep[e.g.,][]{Noeske2007,Daddi2007,Elbaz2011,Whitaker2012,Schreiber2015}. It is thought that this tight correlation is created by the presence of slow and continuous inflows of cold gas streams from cosmic filaments that feed star formation in these galaxies \citep[e.g.,][]{Dekel2009,Ciotti2010}. Mergers, on the other hand, can have much larger star formation rates for their stellar mass. Galaxy merger events introduce and disturb large volumes of gas, driving it inwards, and enabling the rapid creation of a large number of stars \citep[e.g.,][]{Hopkins2009,Veilleux2009}. While the supply of gas available in both of these modes can be driven inwards to the AGN through the loss of angular momentum via gravitational torque processes \citep{Garcia-Burillo2005}, it is still not apparent that these two modes lead to correlated SMBH growth and star formation (see \citealt{Alexander2012}, for a review).

Therefore, much recent work has focused on studying the link, or lack thereof, between star formation and SMBH growth. Several studies have found that the star formation rate (SFR) of a galaxy is correlated with its AGN's luminosity \citep[e.g][]{Mullaney2012,Chen2013,Delvecchio2015,Harris2016,Lanzuisi2017}, while others have found either a shallow SFR to AGN luminosity correlation or no correlation \citep[e.g.,][]{Azadi2015,Stanley2015,Stanley2017,Shimizu2017}. A few works examine the relation in bins of AGN luminosity and find that the relation is luminosity and redshift dependent, with only higher luminosity AGN ($L_{\rm AGN}>10^{44}$ erg s$^{-1}$) and lower redshift ($z<1$) galaxies exhibiting a steep correlation, while lower luminosities or higher redshifts produce flattened relations \citep[e.g.,][]{Shao2010,Rosario2012,Harrison2012}; these findings were also supported by later semi-analytic work by \cite{Gutcke2015}.

Further theoretical work found that the disagreement could arise from the method of analysis used; specifically, that the bivariate distribution of SFR and AGN luminosity gives differing results dependent on the whether the data is binned by SFR or AGN luminosity \citep{Volonteri2015b,Volonteri2015a}. Earlier work by \cite{Hickox2014} found that this disagreement could be caused by the differences in timescales between the two processes, with measurable SFR being averaged over $\sim$100 Myr while AGN X-ray luminosity varies on much shorter timescales. Therefore using AGN luminosity measurements taken at one point in time would introduce scatter.

Another prediction of \cite{Volonteri2015a} is that SMBH growth is better correlated with nuclear SFR ($r<5$ kpc), and that a relation between AGN luminosity and SFR weakens or disappears for SFR integrated over larger areas, i.e. ``global" SFR. This idea was supported by earlier observations by \cite{Diamond-Stanic2012}.

Still other works focus less on the direct connection between SMBH growth and star formation, but instead compare AGN host galaxies to general samples of star-forming galaxies. Some of these find that AGN host galaxies tend to lie primarily along the star-forming main sequence \citep[e.g.,][]{Rosario2013,Stanley2017}, while others find that they tend to lie below the main sequence, having a lower SFR on average for a fixed galaxy mass \citep[e.g.,][]{Shimizu2015,Mullaney2015}. This points to an uncertainty as to whether AGNs are primarily found in quiescent or star-forming galaxies and, importantly, whether star formation and AGNs are similarly triggered.

This work addresses some of these lingering observational conflicts. By using a systematic AGN selection method and deriving all AGN and host galaxy properties from spectral energy distributions (SEDs), we create the Advanced Camera for Surveys Active Galactic Nuclei (ACS-AGN) Catalog. This is a catalog of 2585 AGN host galaxies imaged by the Advanced Camera for Suveys (ACS) on the \textit{Hubble Space Telescope} (\textit{HST}) with redshifts of $0.2<z<2.5$, along with uniformly derived AGN and galaxy properties. Typically, AGN and star formation contributions can be difficult to disentangle, but we avoid this problem by fitting AGN and galaxy SED components simultaneously. This approach also enables us to average out AGN variability and decrease its effect on AGN luminosity measurements due to the broadband nature of our photometric input. Further, by deriving all properties from the same SED fit, we ensure that they are self-consistent. 

Here we present and make publicly available the ACS-AGN Catalog along with analysis examining the relation between AGN host galaxies and the star formation main sequence and the relation between AGN luminosity and SFR. The paper is organized as follows: Section \ref{sec: ACS-GC} presents our initial sample of galaxies; Section \ref{sec:AGN select} discusses the AGN selection; in Section \ref{sec: Galaxy SEDs} we discuss the fitting of SED components to our AGN host galaxies using multi-band photometric data; in Section \ref{sec: gal props} we discuss the derivation of our AGN and host galaxy properties; in Section \ref{sec: results} we present our findings from statistical analysis of the AGN and host galaxy properties; and finally, in Section \ref{sec: conclusions} we discuss our conclusions. Throughout this paper, we use the Planck 2015 cosmology of $H_{0} = 67.8$ km s$^{-1}$ Mpc$^{-1}$, $\Omega_{M}=0.308$, and $\Omega_{\Lambda}=0.692$ \citep{Planck2016}.

\section{Parent Galaxy Sample} \label{sec: ACS-GC}
Our input galaxy sample was the Advanced Camera for Surveys General Catalog \citep[ACS-GC;][]{Griffith2012}. The ACS-GC is a photometric and morphological catalog of 469,501 galaxies imaged by the ACS on \textit{HST} in four surveys: the Galaxy Evolution from Morphologies and SEDs (GEMS) survey, the Cosmological Evolutionary Survey (COSMOS), the Great Observatories Origins Deep Survey (GOODS), and the All-wavelength Extended Groth Strip International Survey (AEGIS). The ACS-GC provides photometric and/or spectroscopic redshifts for 345,783 (74\%) of its galaxies, spanning a redshift range of $z\lesssim6$ with a median redshift of $\langle z \rangle=0.885$. This subsample of ACS-GC galaxies with redshifts is our parent galaxy sample.

The GEMS survey \citep{Caldwell2008} uses the F606W and F850LP filters, has a coverage area of 0.21 deg$^{2}$, and a 5$\sigma$ limiting AB magnitude of 25.7. COSMOS \citep{Scoville2007} uses the F814W filter, has a coverage area of 1.8 deg$^{2}$, and a 5$\sigma$ limiting AB magnitude of 26.0. GOODS \citep{Dickinson2003} uses the F606W, F775W, and F850LP filters, has a coverage area of 0.14 deg$^{2}$, and a 5$\sigma$ limiting AB magnitude of 25.7. The GOODS survey is split into the GOODS-N (North) field and GOODS-S (South) field, with the GEMS field enveloping the GOODS-S field. Lastly, AEGIS \citep{Davis2007} uses the F606W and F850LP filters, has a coverage area of 0.20 deg$^{2}$, and a 5$\sigma$ limiting AB magnitude of 26.2. The ACS-GC is dominated by COSMOS galaxies due to its large field; COSMOS galaxies are 65\% of all ACS-GC galaxies while GEMS + GOODS-S, AEGIS, and GOODS-N galaxies constitute 15\%, 15\%, and 5\% of ACS-GC galaxies, respectively.

\section{Active Galaxy Selection} \label{sec:AGN select}
Infrared (IR) and X-ray data are commonly used to identify galaxies containing an AGN \citep[e.g.][]{Stern2005,Brusa2010}. During AGN accretion, gas is accreted and heated in the accretion disk, emitting strongly in the ultraviolet (UV). Some of these UV photons can interact with the AGN's hot compact plasma corona and undergo inverse Compton scattering, shifting them into the X-ray regime. This makes AGNs an ideal target for detection in the X-ray if they are unobscured, or even moderately obscured \citep{Brandt2015}. A common criterion for selecting AGNs in the X-ray regime is a rest-frame luminosity cut in the 2 -- 10 keV band \citep{Brandt2015}. In addition to interacting with the corona, some of the UV light from the AGN accretion disk is reprocessed into lower energy wavelengths as it heats nearby material, e.g. a torus (see \cite{Brandt2015} for an in-depth review of AGN physics and selection techniques). Some IR AGN selection techniques focus on the mid-IR regime and compare flux in different bands (color cuts) in the mid-IR, selecting for the spectral signature of an AGN torus-like structure that has been heated by its host AGN \citep[e.g.][]{Lacy2004,Stern2005,Donley2012}. The mid-IR emission from the AGN torus is largely unattenuated by dust, making IR AGN selection complementary to X-ray AGN selection \citep{Assef2013}. The most common color cuts used to select AGNs in the mid-IR are from \cite{Lacy2004,Lacy2007}, \cite{Stern2005, Stern2012}, and \cite{Donley2012}. 

In this Section we create our active galaxy sample using IR and X-ray AGN selection criteria, finding 3955 AGN host galaxies. In Section \ref{subsec:IR} we select AGNs using data from the \textit{Spitzer Space Telescope}. In Section \ref{subsec:XRAY} we select AGNs using data from the \textit{Chandra X-ray Observatory}. In Section \ref{subsec:Comparison} we compare the different AGN samples obtained from each telescope.

\subsection{Infrared AGN Selection} \label{subsec:IR}
In the IR regime, we created our active galaxy sample using data from the Infrared Array Camera (IRAC) on \textit{Spitzer} \citep{Fazio2004}. IRAC has observed all four of the ACS-GC \textit{HST} fields in four bands, channel 1: 3.6 $\mu$m, channel 2: 4.5 $\mu$m, channel 3: 5.8 $\mu$m, and channel 4: 8.0 $\mu$m, with a resolution of $\sim$2\farcs5 for all channels.

We conducted a spatial crossmatch between the ACS-GC and the NASA/IPAC Infrared Science Archive (IRSA)\footnote{\url{http://irsa.ipac.caltech.edu}} source catalog for \textit{Spitzer} (the \textit{Spitzer} Enhanced Imaging Products (SEIP) Source List), using the IRSA Gator search tool. We used an initial conical search radius of 3$\farcs$0 for each source in order to fully understand the distribution of spatial separations between sources from the ACS-GC and the SEIP. We then found the $3\sigma$ radius of this distribution of source separations. This $3\sigma$ radius should encompass nearly all true matches (99.7\% if Gaussian) while minimizing false matches between sources from each catalog. We found that the 3$\sigma$ radius of the SEIP crossmatches in these fields was 0$\farcs$5. We redid the crossmatch, requiring the source separations to be less than or equal to this value.

The matched source separations were approximately Gaussian distributed, as expected from small random differences in position between the optical and IR sources and/or statistical fluctuations in measured source positions. The distributions also had small systematic offsets for each extragalactic field (i.e., GEMS + GOODS-S, COSMOS, GOODS-N, AEGIS). We calculated these systematic offsets by fitting a Gaussian profile to each distribution of source separations, and present the astrometric offsets for each field in Table \ref{tab: offsets}.

\begin{deluxetable}{llllll}[t!]
	\tablewidth{0pt}
	\tablecolumns{4}
	\tablecaption{Systematic offsets in source matching\label{tab: offsets}}
	\tablehead{
		\colhead{ACS-GC} &
		\colhead{IR / X-ray} &
		\colhead{RA Offset} &
		\colhead{Dec Offset} &  \\ 
		\hspace{.4cm}Field  & \hspace{.3cm}Catalog  & \hspace{.6cm}[arcsec] & \hspace{.6cm}[arcsec]
	}
	\startdata
	GEMS + & SEIP & $+0.058\pm0.001$ & $-0.213\pm0.002$ \\
	GOODS-S & CSC & $-0.070\pm0.016$ & $+0.282\pm0.024$ \\
	COSMOS & SEIP & $-0.053\pm0.001$ & $+0.092\pm0.001$ \\
	& CSC & $+0.027\pm0.006$ & $+0.032\pm0.007$ \\
	GOODS-N & SEIP & $-0.009\pm0.003$ & $-0.288\pm0.005$ \\
	& CSC & $+0.068\pm0.015$ & $-0.118\pm0.023$ \\
	AEGIS & SEIP & $-0.029\pm0.002$ & $-0.200\pm0.002$ \\ 
	& CSC & $-0.082\pm0.011$ & $+0.156\pm0.117$
	\enddata
	\tablecomments{Systematic offsets based on Gaussian fits to positional differences between the ACS-GC and SEIP/CSC catalogs for each field (offset $\equiv$ SEIP/CSC $-$ ACS-GC). \vspace*{-1cm}}
\end{deluxetable}

We then adjusted the \textit{HST} coordinates for these systematic offsets and did a final crossmatch, using a radius of 0$\farcs$5 for SEIP sources. We retain the original ACS-GC coordinates in our catalog. Correcting for the systematic astrometric offsets increased the number of matched sources within our search cone by 2.7\% for SEIP sources.

We then identified galaxies hosting AGNs using the established \cite{Donley2012} IR AGN selection criteria. Implicit with these criteria, each source was required to have data in all four IRAC channels; approximately 43\% of matched \textit{Spitzer} sources met this requirement. Most incompleteness was in the 5.8 $\mu$m and 8.0 $\mu$m bands, which are approximately an order of magnitude less sensitive than the 3.6 $\mu$m and 4.5 $\mu$m bands of \textit{Spitzer}. Our selection resulted in 1861 IR-selected AGN host galaxies (2.3\% of all matched \textit{Spitzer} sources with four channel flux data).

\subsection{X-Ray AGN Selection} \label{subsec:XRAY}
To create our active galaxy selection in the X-ray regime, we used data from the Advanced CCD Imaging Spectrometer (ACIS) on \textit{Chandra}, presented in the \textit{Chandra} Source Catalog (CSC) v2.0 \citep{Evans2018} preliminary detection list (pd1).

We repeated our aforementioned crossmatching procedure, conducting an initial spatial crossmatch between the ACS-GC catalog and the CSC 2.0 pd1 using a 2$\farcs$0 matching radius, finding the $3\sigma$ radius of the resulting distribution (1$\farcs$5), and redoing the crossmatch using this value. We then repeated our procedure of Gaussian fitting and removal of systematic offsets in the positional differences between matched sources as outlined in Section \ref{subsec:IR}; the offsets for each field can be found in Table \ref{tab: offsets}. Correcting the systematic astrometric offsets increased the number of matched sources within our \textit{Chandra} search cone by 0.21\%.

The AGN selection criterion we used for the X-ray regime was a rest-frame X-ray luminosity cut in the 2 -- 10 keV band of $L_{2-10}>10^{42}$ erg s$^{-1}$; this criterion ensures the exclusion of all but the most vigorously star-forming galaxies -- those with SFR $\geq$ 200 $M_{\odot}$ yr$^{-1}$ \citep{Ranalli2003}. In order to obtain a rest-frame luminosity in the 2 -- 10 keV band, we first converted the observed 2 -- 7 keV photon flux from the CSC to energy flux using an effective energy\footnote{\url{http://cxc.harvard.edu/csc/columns/ebands.html}} of 3.8 keV. We then assumed a power law X-ray spectrum with a photon index of 1.7, the mean value found for a comparable AGN sample by \cite{Brightman2014}, in order to model the observed 2 -- 10 keV energy flux. Finally, we used galaxy redshift data from the ACS-GC catalog in order to apply a K-correction to the observed 2 -- 10 keV energy flux and calculate the rest-frame 2 -- 10 keV luminosity for each object. We then selected AGN as galaxies that had restframe 2 -- 10 keV band luminosity 1$\sigma$ lower bounds greater than our luminosity cutoff of $10^{42}$ erg s$^{-1}$.

Our selection resulted in 2624 X-ray-selected AGN host galaxies (83.5\% of all matched \textit{Chandra} sources with flux data in the 2 -- 10 keV band), with a significant number of these also selected as AGN in the IR. This overlap is discussed further in Section \ref{subsec:Comparison}.

\subsection{AGN Selection Comparison} \label{subsec:Comparison}
X-ray and IR selection techniques uncover different samples of AGNs that have some overlap. \cite{Eckart2010} found that X-ray AGNs that lack high-ionization and/or broad lines in the optical were less likely to be selected as IR AGNs, and further suggest that IR-selected AGNs that are not selected in the X-ray are primarily high-luminosity AGNs that are obscured and/or at high redshift. Based on a sample of 55 AGNs in the COSMOS, GOODS-N, and EGS fields, \cite{Azadi2017} find many selection biases in X-ray and IR-selected AGNs, including a bias towards high-mass galaxies in X-ray-selected AGNs and a bias towards moderate-mass galaxies in IR-selected AGNs. They attribute this to the fact that X-ray selection techniques can identify AGNs at low specific accretion rates, while IR selection techniques are biased towards finding high specific accretion rate AGNs. \cite{Azadi2017} also finds that IR-selected AGNs are biased towards lower dust content than X-ray-selected galaxies, which they attribute to the stellar mass selection bias of IR-selected AGNs. Specifically, IR-selected AGNs are found in lower stellar mass galaxies, which tend to be dustier than higher mass galaxies. Further, \cite{Hainline2016} found that an IR-selected AGNs sample can be contaminated by star-forming dwarf galaxies that are capable of mimicking the IR colors of more luminous AGNs.

\begin{figure}[!t]
	\begin{center}
		\includegraphics[width=0.45\textwidth]{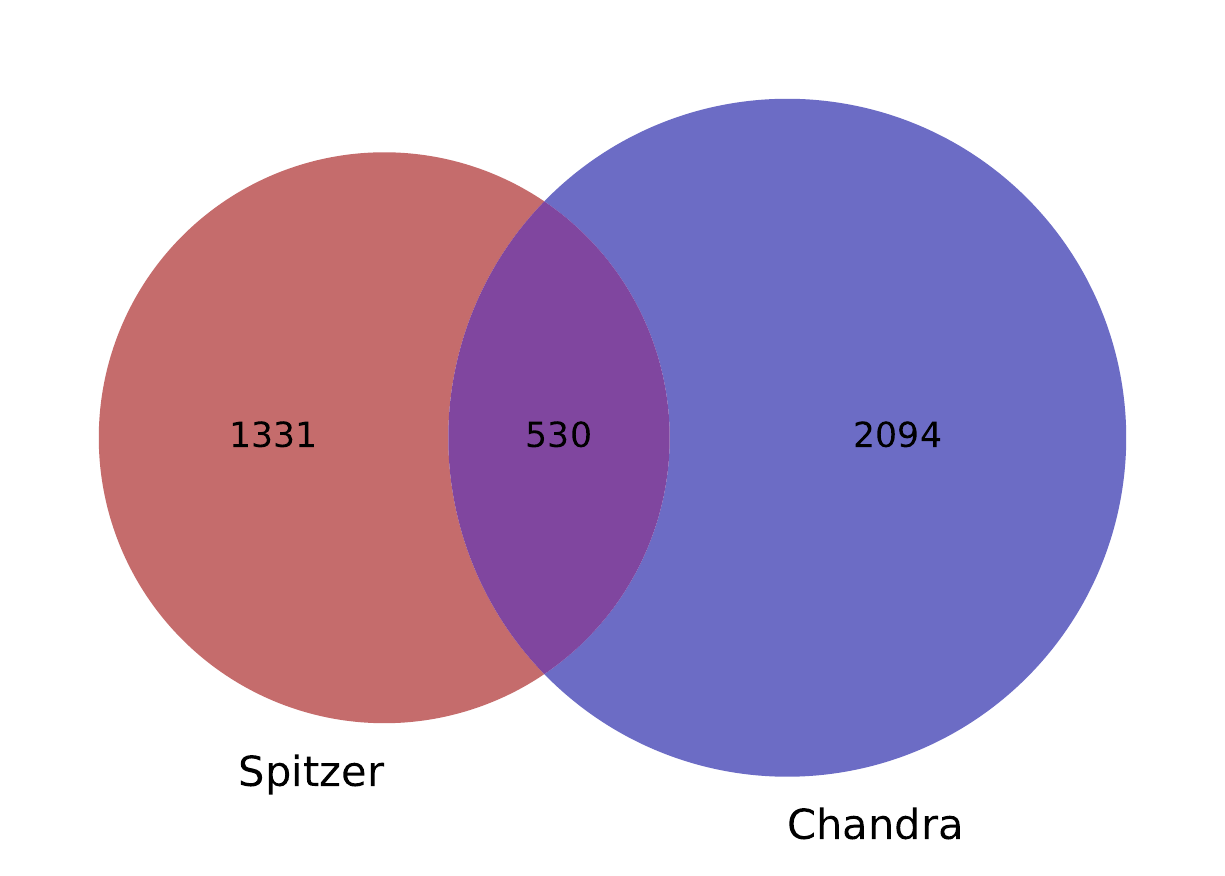}
	\end{center}
	\vspace{-12pt}
	\caption{Venn diagram of identified AGN host galaxies: IR-selection by \textit{Spitzer} and X-ray-selection by \textit{Chandra}.}\label{plot: venn}
\end{figure}

Through our selection process, we narrowed our initial sample of 345,783 ACS-GC galaxies with redshifts to 3955 unique active galaxies identified by IR and/or X-ray methods. Of these, 2094 and 1331 were uniquely identified by \textit{Chandra} and \textit{Spitzer}, respectively, while 530 were identified by both telescopes (Figure \ref{plot: venn}). With the wide redshift range and galaxy parameter space covered by the deep \textit{Spitzer} and \textit{Chandra} observations in these fields, we see the benefit of using X-ray and IR selection techniques in tandem as their biases work to complement each other.

Examining the COSMOS field --- our largest field, with an area of $\sim$1.6 deg$^{2}$ --- we find number densities of 1154 X-ray-selected AGNs per deg$^{2}$ and 818 IR-selected AGNs per deg$^{2}$. We can compare these numbers to those of \cite{Mendez2013}, a study that compares X-ray and IR AGN selection techniques in the COSMOS field. We find the number density of X-ray-selected AGNs to be in good agreement (1154 vs 1176), while our IR-selected AGNs number density disagrees significantly (818 vs 443). This may be due to the increase in \textit{Spitzer} observation depth since that work, with \cite{Mendez2013} reporting flux limits of $\gtrsim$20 $\mu$Jy in the 5.8 $\mu$m channel while the SEIP data typically have flux densities (3$\sigma$+) as low as 0.01-1 $\mu$Jy in the 5.8 $\mu$m channel.

\section{Obtaining Galaxy SEDs} \label{sec: Galaxy SEDs}
We use SEDs in order to create self-consistent models based on photometric data to derive AGN and galaxy properties. Since observations that span a large portion of a galaxy's spectrum are difficult to obtain, fitting SED templates to available photometric data is often necessary. The Low-Resolution Templates (\texttt{LRT}) program \citep{Assef2008,Assef2010} is one such template fitting package that fits galaxy and AGN templates at the same time.

In this Section we model our AGN host galaxies' SEDs and employ Monte Carlo technique error modeling, resulting in 2873 AGN host galaxies with well fit SEDs. In Section \ref{subsec: photo dat} we discuss the multi-band photometric data we obtained and used to model our SEDs. In Section \ref{subsec: SED fitting} we discuss how we fit galaxy and AGN SED templates to the multi-band photometric data of our active galaxies. Finally, in Section \ref{subsec: MC error} we discuss the method by which we estimated errors on our SED-derived galaxy parameters through the use of Monte Carlo techniques.

\subsection{Photometric Data}\label{subsec: photo dat}
We first obtained multi-band photometric data for our AGN host galaxies, requiring flux data in at least seven bands for each galaxy. This is because \texttt{LRT} requires flux data in at least seven photometric bands from 0.03 $\mu$m -- 30 $\mu$m in order to properly fit SED templates and an extinction parameter and maintain at least 1 degree of freedom in the model. 

For most galaxies in the COSMOS field, we used the COSMOS2015 catalog from \cite{Laigle2016}. This catalog includes flux and/or magnitude data from the X-ray to radio; however, we only used data in 22 of the bands, from the mid-UV to the mid-IR, due to the 0.03 -- 30 $\mu$m wavelength range of our SED fitting software. These data came from instruments on \textit{GALEX}, Subaru, CFHT, VISTA, and \textit{Spitzer}. Of the 2639 galaxies we selected as active in the COSMOS field, the COSMOS2015 catalog contained at least seven bands of photometric data for 2587 (98.0\%) of them.

Unlike the COSMOS survey, the other three surveys included in this work do not have large, complete, multi-band photometric catalogs publicly available. For the remaining 1316 non-COSMOS galaxies we selected as active, we obtained photometry for 576 (43.7\%) of them in 18 bands, ranging from the near-UV to the mid-IR. These data came from the CFHT and VISTA telescopes (data provided by Dan Masters, private correspondence), as well as \textit{WISE} and \textit{Spitzer} (data from the AllWISE Source Catalog and the SEIP Source List). In total, we obtained at least seven bands of photometric data for 3163 of the 3955 (79.9\%) galaxies we selected as active.

\subsection{SED Fitting}\label{subsec: SED fitting}
We use the multi-band photometric data to fit SED templates to our active galaxies with the \texttt{LRT} software. \texttt{LRT} models a galaxy's SED as a non-negative linear combination of an AGN and three galaxy SED components: elliptical, SBc spiral, and irregular. The AGN SED component is a Type 1 AGN to which an extinction law is applied \citep{Cardelli1989,Gordon1998}. The extinction term $E_{B-V}$ is fit by \texttt{LRT} and mimics nuclear obscuration when applied, allowing for the AGN template to resemble an obscured AGN. The low-resolution templates were empirically derived from over 16,000 galaxies in the Bo\"{o}tes field with spectroscopic redshifts and photometry, and are limited to the wavelength range of 0.03 -- 30 $\mu$m \citep{Assef2008,Assef2010}. These templates are simultaneously fit to the observed photometric data points, accounting for the associated error on the flux measurements.

The creation of empirically derived templates involves condensing a large dataset with intrinsic scatter into a singular SED template. This process creates uncertainty in the template itself, and may be one of the largest sources of error in SED modeling \citep{Abrahamse2011}. In order to account for template uncertainty, we instituted an error floor of 10\% on all of our photometric data (i.e., if the associated error on a given flux measurement was $<$10\%, we fixed it to 10\%). This practice is also used by other studies to account for known errors \citep[e.g. instrument calibration errors,][]{Donley2012}.

In addition, the number of inaccurate photometric data points increases as the size of the data set increases. Anomalous photometric data can come from multiple sources, encompassing both instrumental and astrophysical errors (e.g. variability, foreground/background). The presence of anomalous data points will lead to poor SED fits. A method used to minimize the impact of anomalous data points is to iteratively exclude individual photometric data points, fit the SED templates, and calculate whether the fit is substantially improved from its parent using the $F$ ratio statistic, which compares the improvement in $\chi^{2}$ with the change in degrees of freedom \citep[e.g.,][]{Chung2014}. We used this method in our SED fitting, allowing up to two photometric data points to be excluded for each galaxy being modeled, as long as there was at least seven data points still being fit, requiring a p-value $\leq0.05$ in the $F$ ratio test, as well as a $\chi^{2}_{\nu}$ goodness of fit cutoff. The model was only found satisfactory if $\chi^{2}_{\nu}<1+\sqrt{2/\nu}$, where $\nu$ is the degrees of freedom ($\nu=$ data points -- fit parameters -- 1) and $\sqrt{2/\nu}$ is the $1\sigma$ error, assuming Gaussian uncertainties \citep{Chung2014}.

\begin{figure*}[t!]
	\centering
	\includegraphics[width=0.47\textwidth]{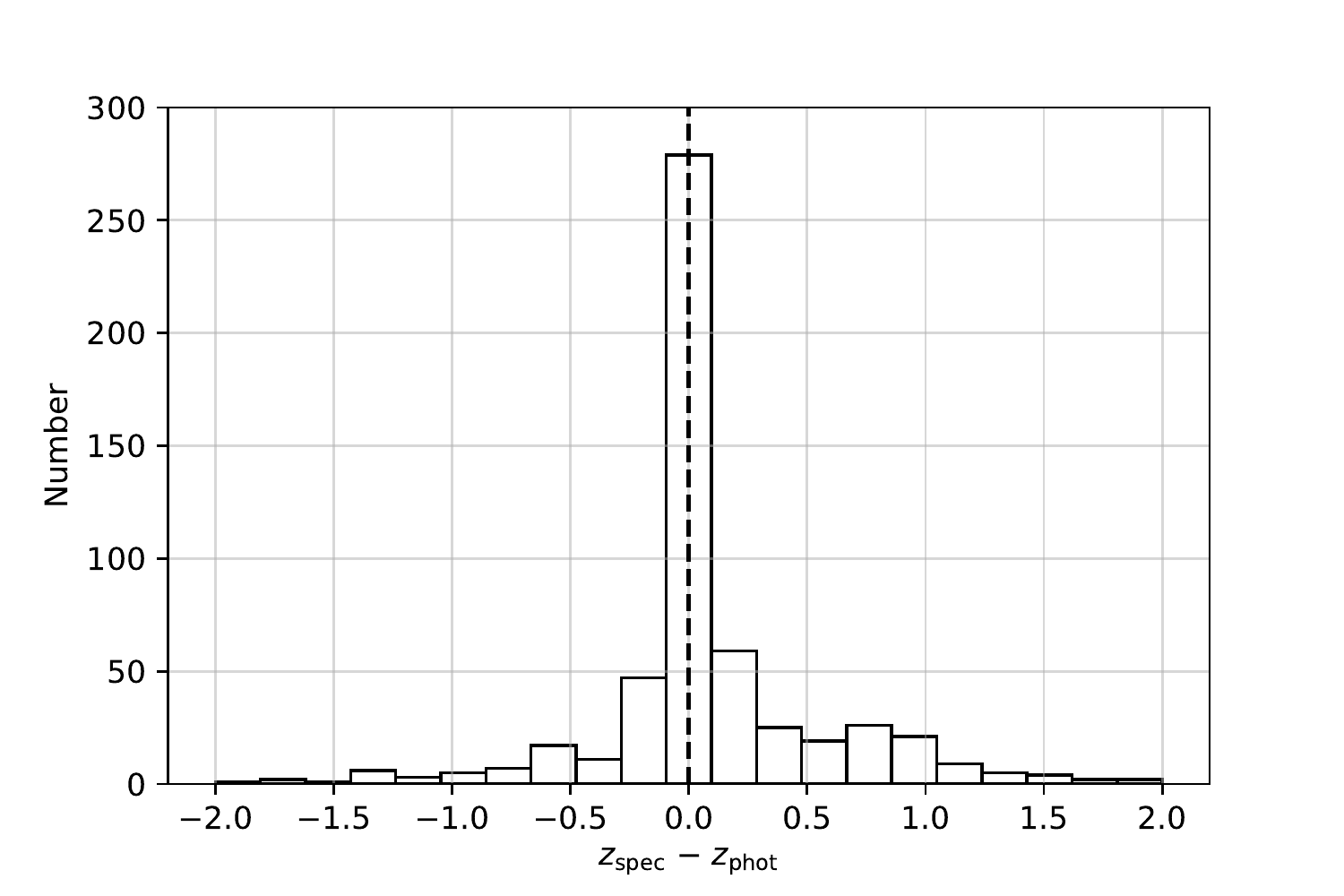}
	\includegraphics[width=0.47\textwidth]{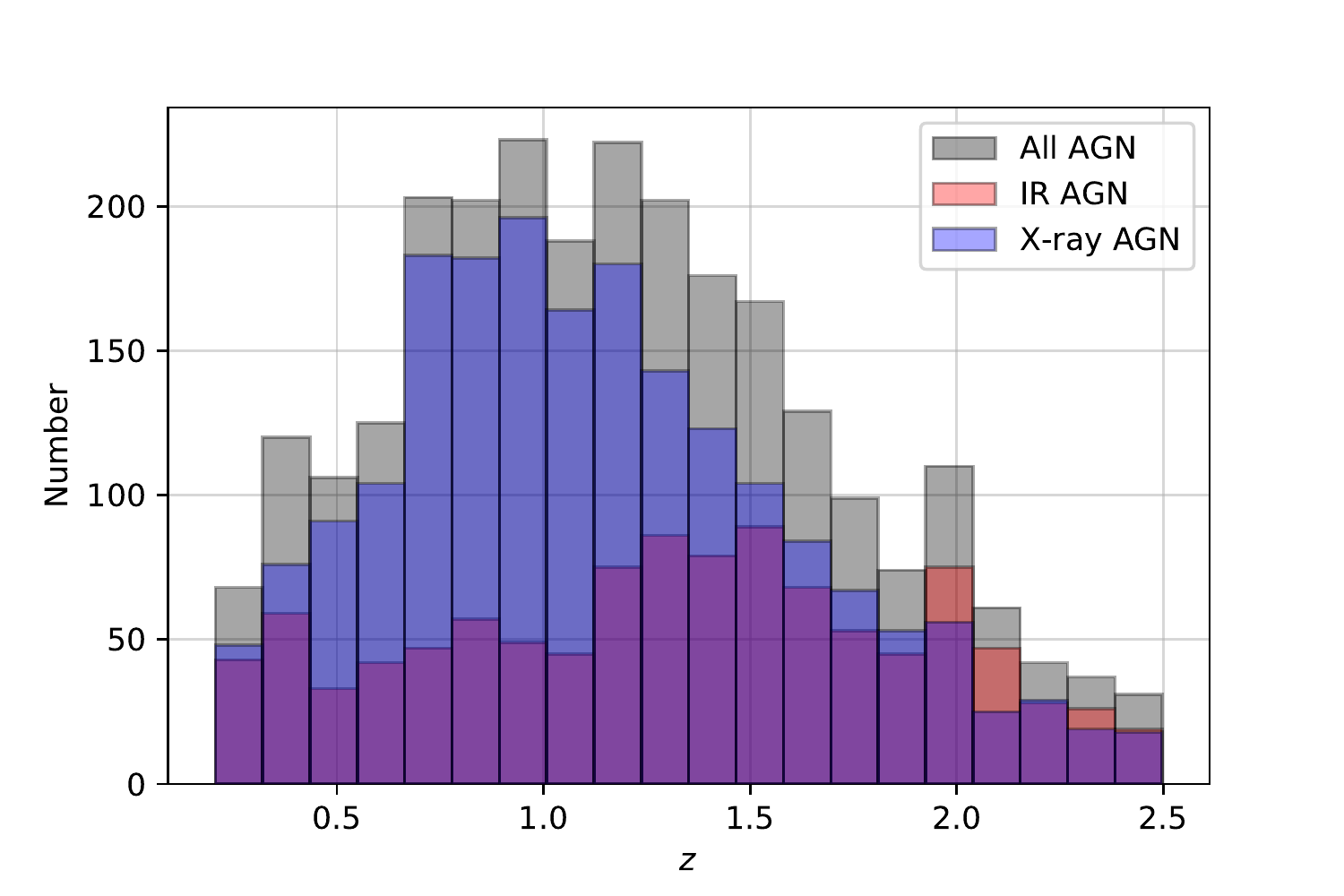}
	
	\caption{Histogram of the differences between spectroscopic and photometric redshift values for the subsample of our galaxies that have both and exist between $0.2<z_{\rm spec}<2.5$ (left); the dashed line is the $z_{\rm spec}=z_{\rm phot}$ line. Histogram of the redshifts for our active galaxy sample (right); note how the X-ray-selected distribution peaks at a lower redshift value than the IR-selected distribution. The All AGN sample (gray) is included in addition to the two subpopulations that make it up (IR AGN [Red] and X-ray AGN [Blue]) because the two selection techniques create an overlapping sample of AGNs.}
	\label{plot: reds}
	\vspace{12pt}
\end{figure*}

\subsection{Monte Carlo Error Estimation}\label{subsec: MC error}
In order to model errors for the AGN and host galaxy properties that we extract from the SED fit (Section \ref{sec: gal props}), we used Monte Carlo techniques. Specifically, after eliminating any anomalous data points (Section \ref{subsec: SED fitting}), we resampled the remaining photometric data, drawing new values from a Gaussian distribution centered on the known value, with standard deviation equal to the associated flux error. After doing this for all data points, we reran \texttt{LRT}, creating a new SED model fit. We repeated this process 1000 times for each galaxy, keeping only those SED model fits that obeyed the goodness of fit criterion outlined in Section \ref{subsec: SED fitting}. We removed from further analysis the galaxies in which fewer than 10 of the 1000 iterations met the goodness of fit criterion.

This process created an SED model fit distribution with up to 1000 individual fits for each AGN host galaxy, propagating the uncertainty from the photometric data to the SED model fits. This allowed us to obtain uncertainty values for the AGN and host galaxy properties derived in the following Section. In total, we were able to satisfactorily model and create SED model fit distributions for 2873 of the 3163 (88.0\%) active galaxies for which we obtained multi-band photometric data. 

\begin{figure*}[t]
	\centering
	\includegraphics[width=0.47\textwidth]{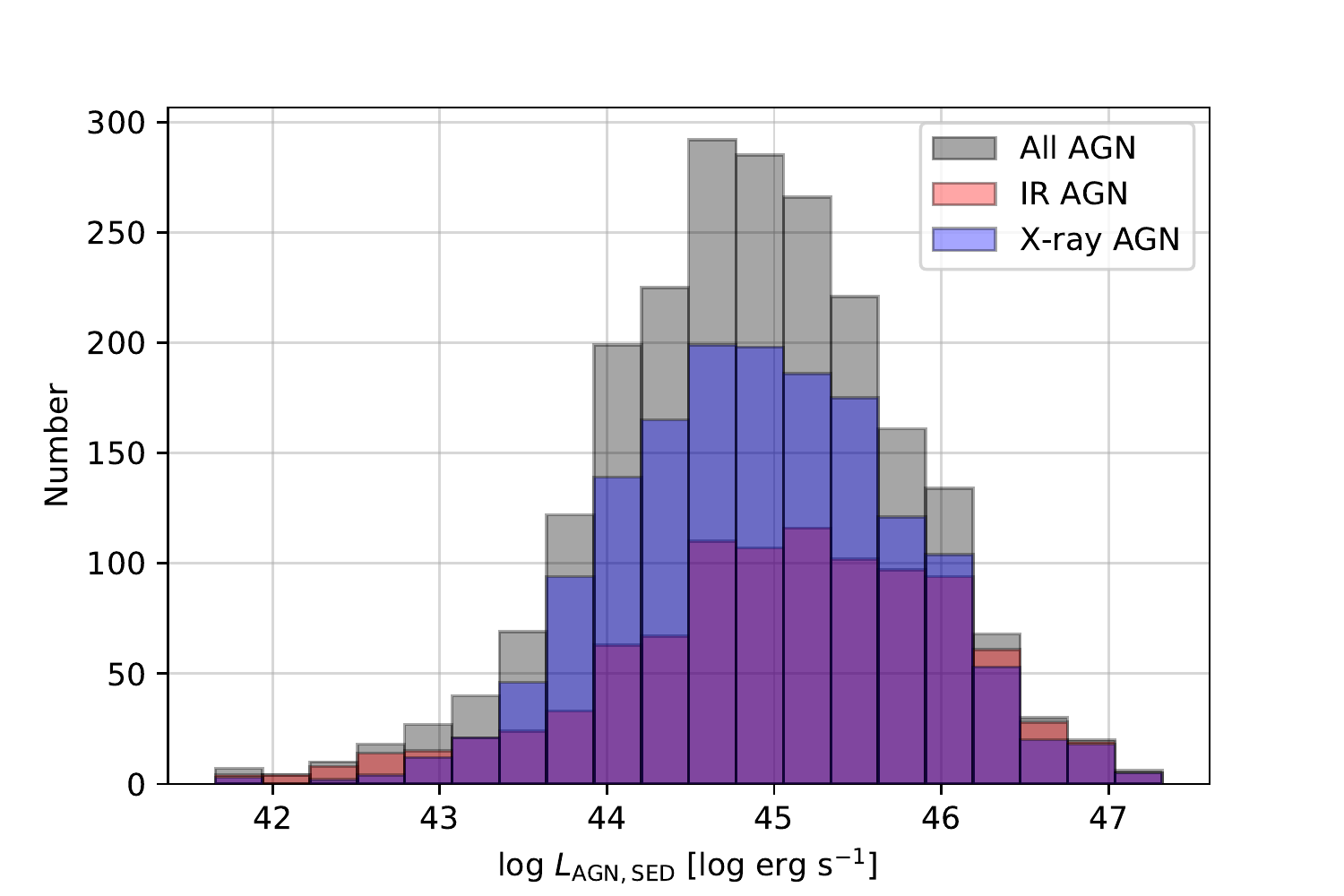}
	\includegraphics[width=0.47\textwidth]{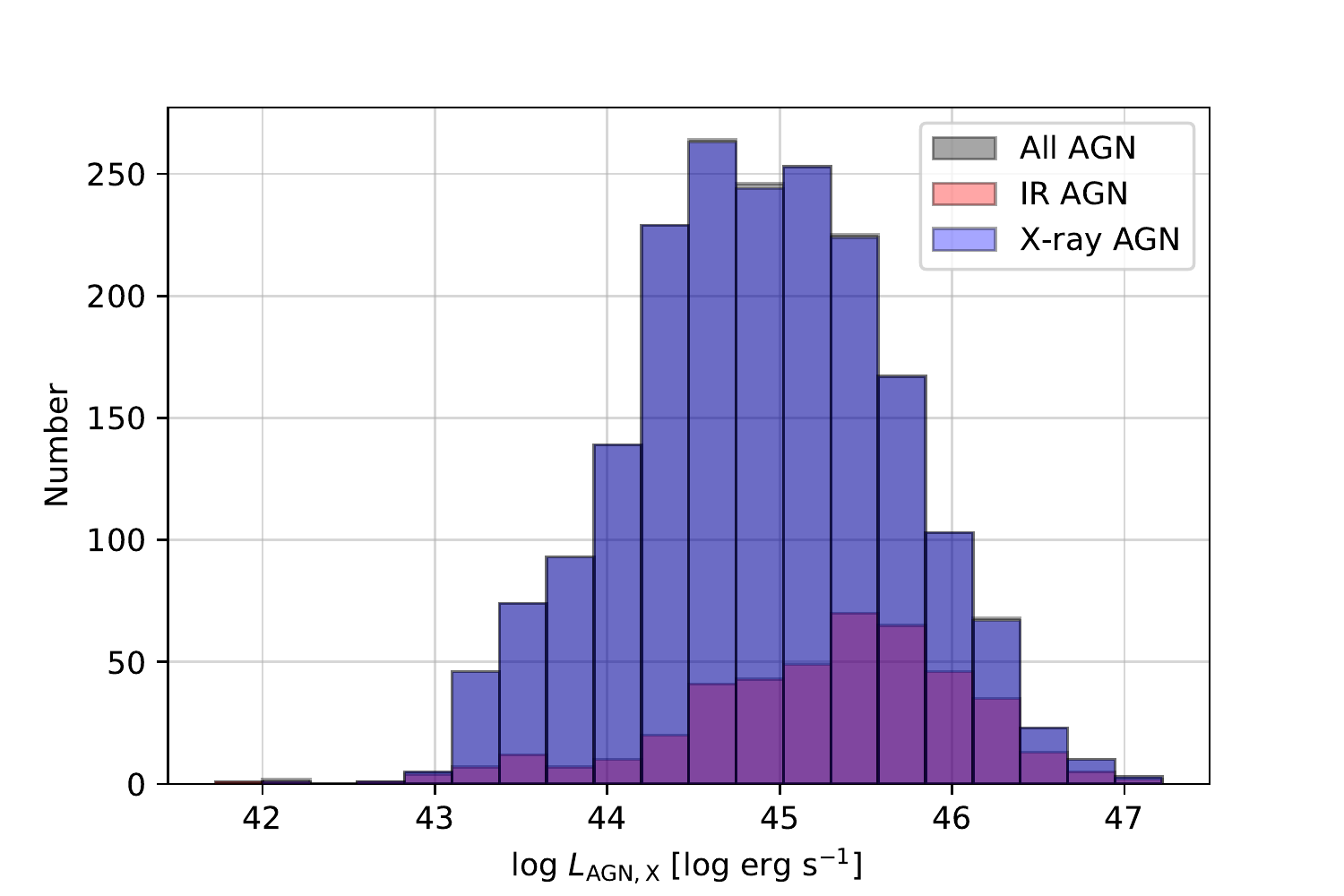}
	\includegraphics[width=0.47\textwidth]{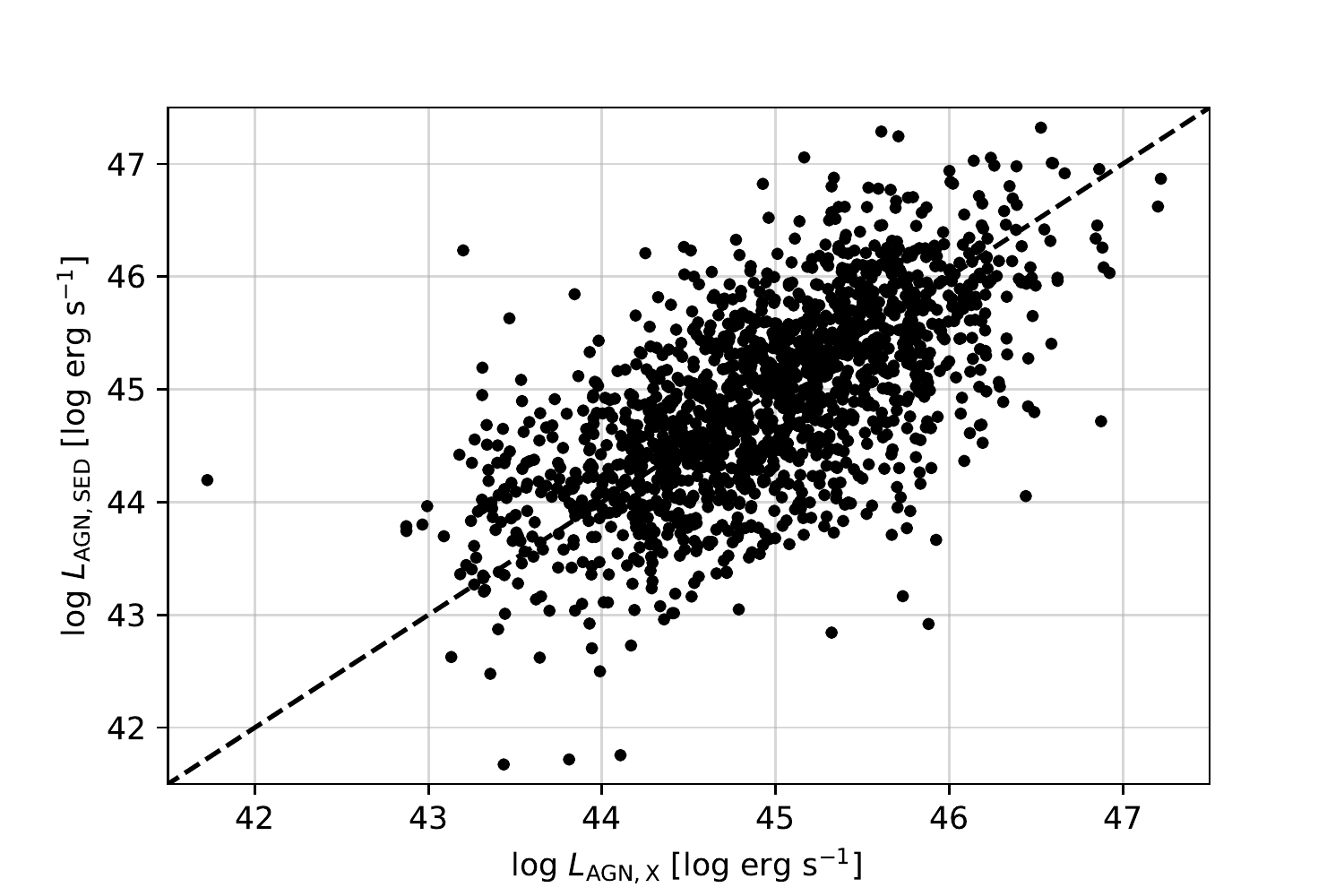}	
	
	\caption{Histogram of the $L_{\rm AGN,SED}$ values for our active galaxy sample (top left); note the similar peaks of the IR and X-ray-selected distributions. Histogram of the $L_{\rm AGN,X}$ values for our X-ray-selected subsample (top right); note the higher luminosity peak for those selected with both IR and X-ray methods. $L_{\rm AGN,SED}$ as a function of $L_{\rm AGN,X}$ for the subsample of  galaxies which have both luminosity measurements (bottom); the dashed line is the $L_{\rm AGN,X}=L_{\rm AGN,SED}$ line. Note that the scatter is evenly distributed across the $L_{\rm AGN,X}=L_{\rm AGN,SED}$ line for all luminosities, and therefore there is no dependence on AGN luminosity.}
	\label{plot: gal props 1}
	\vspace{12pt}
\end{figure*}

\begin{figure*}[!]
	\centering
	\includegraphics[width=0.47\textwidth]{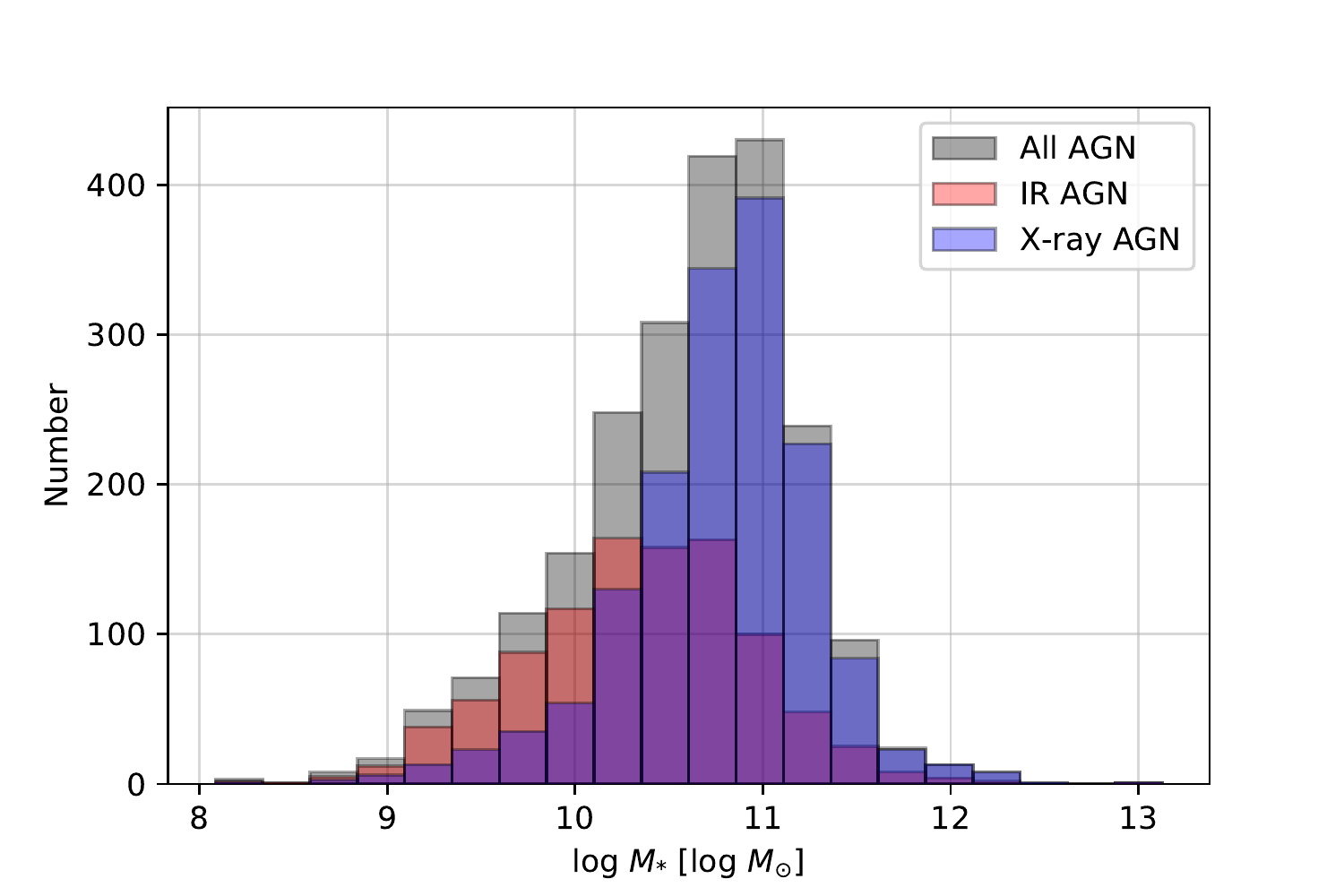}
	\includegraphics[width=0.47\textwidth]{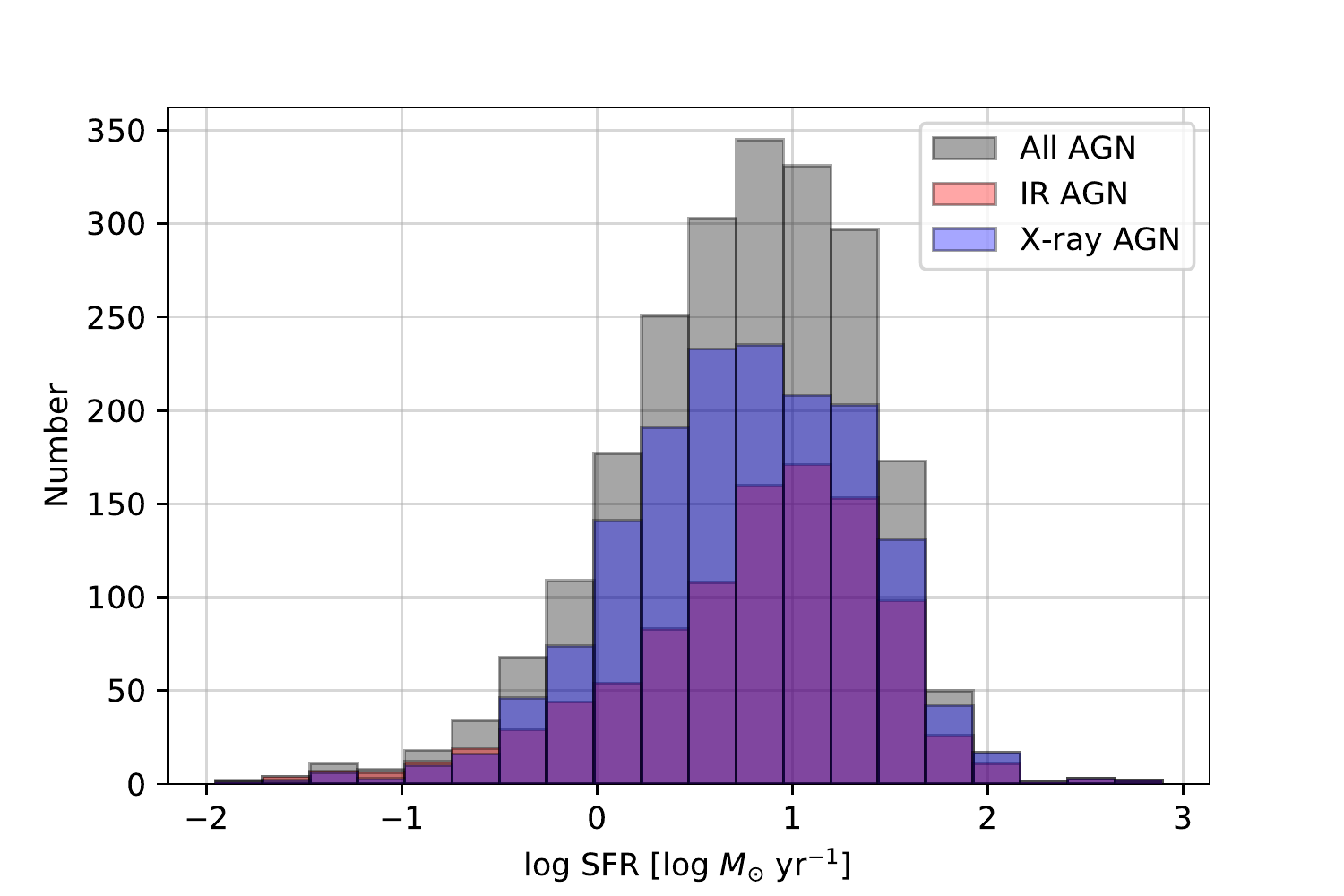}
	\includegraphics[width=0.47\textwidth]{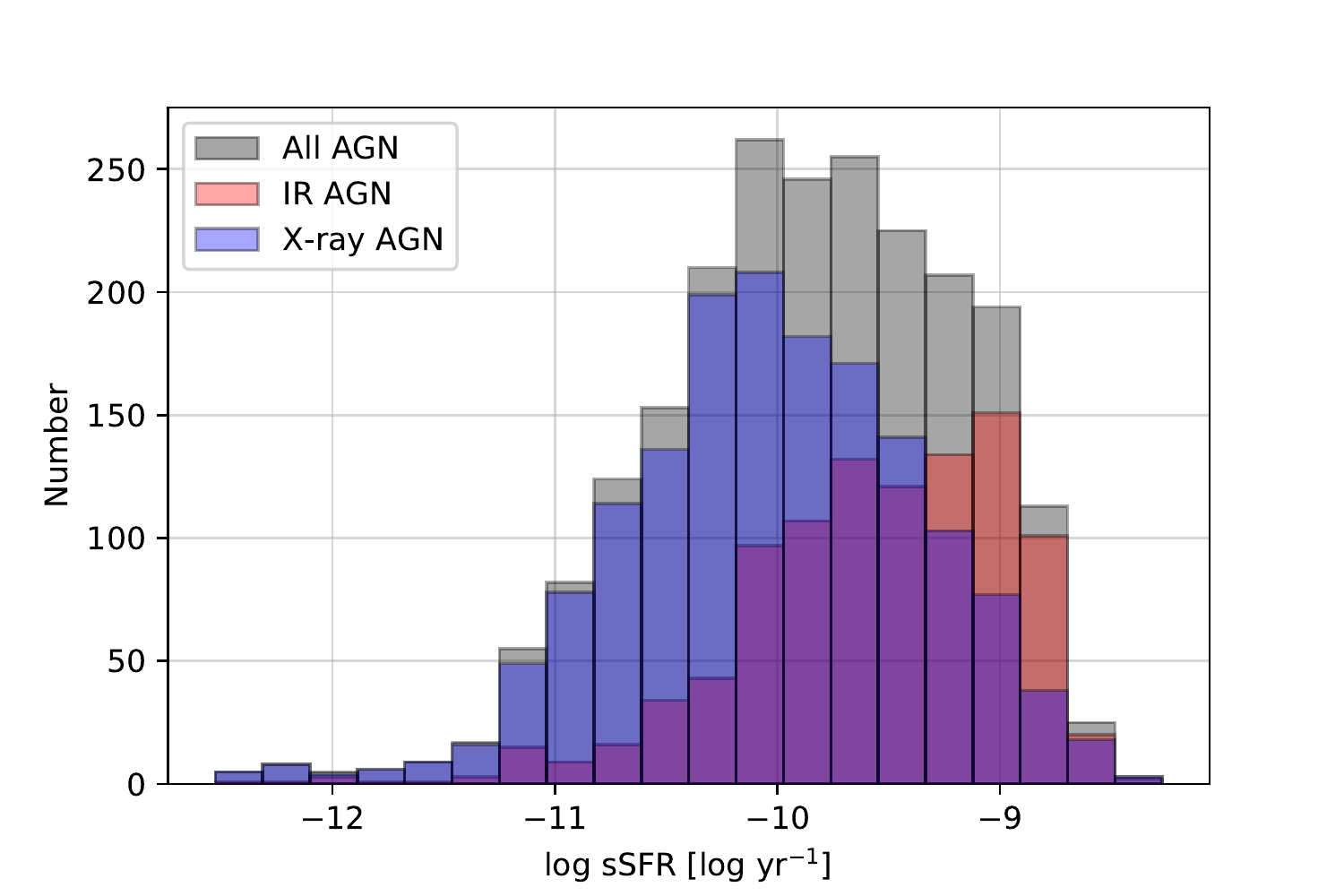}
	\includegraphics[width=0.47\textwidth]{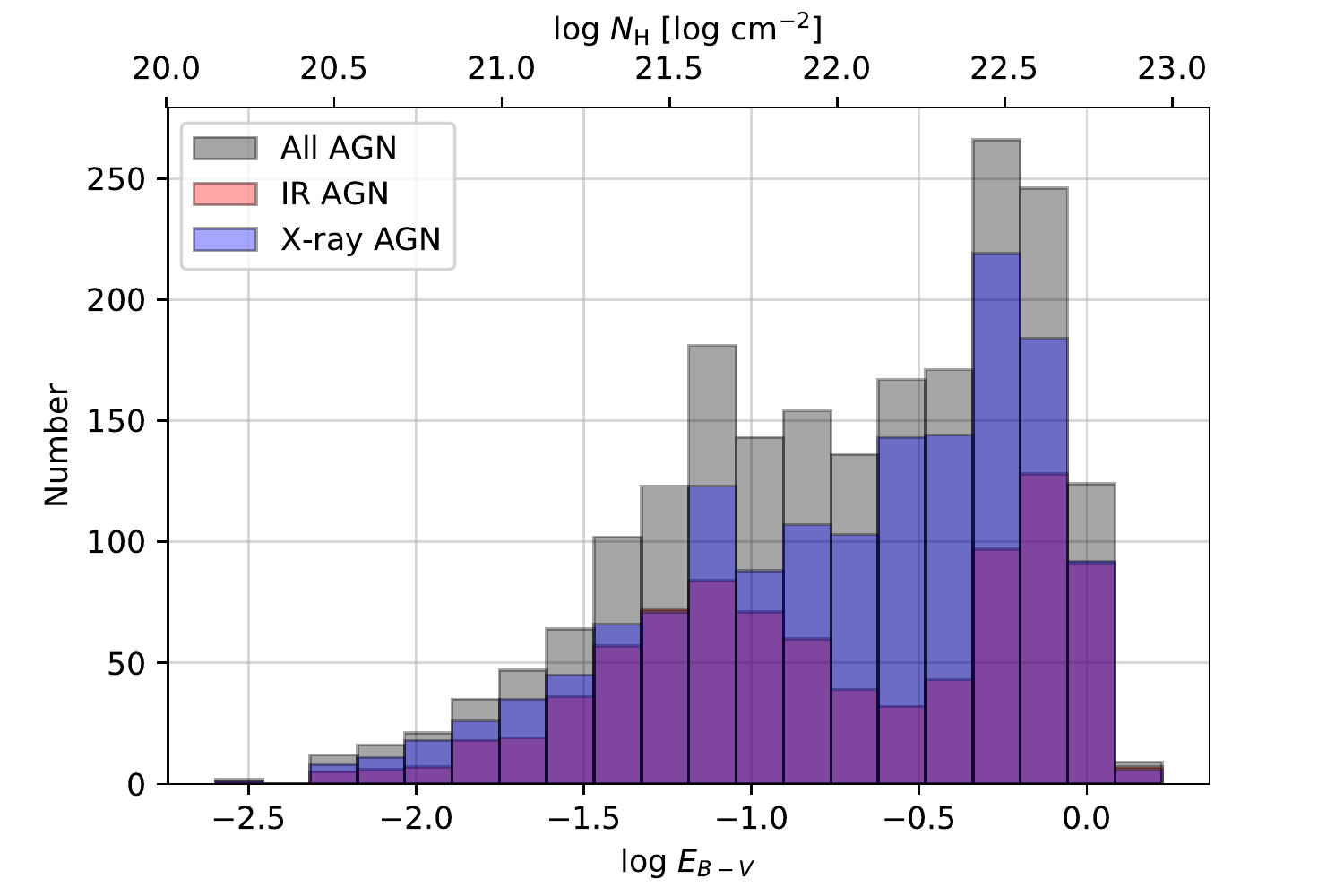}
	
	\caption{Histogram of the $M_{*}$ values for our active galaxy sample (top left), histogram of the SFR for our active galaxy sample (top right), and histogram of the sSFR values for our active galaxy sample (bottom left); note the lower peak $M_{*}$ value of the IR-selected AGN host galaxy distribution compared to the X-ray-selected AGN host galaxy distribution, and the slightly higher SFR peak of the IR-selected AGN host galaxy distribution compared to that of the X-ray-selected AGN host galaxy distribution. These effects combine to produce the two subsamples' sSFR distributions, with the IR-selected AGN host galaxy distribution peaking at a higher value than the X-ray-selected AGN host galaxy distribution. Histogram of the $E_{B-V}$ and $N_{\rm H}$ values for our active galaxy sample (bottom right); note the double peaked nature of the distributions.}
	\label{plot: gal props 2}
	\vspace{12pt}
\end{figure*}

\section{Galaxy Properties} \label{sec: gal props}
All of the AGN and host galaxy properties we use in our analysis, save for the redshifts included in the ACS-GC, are derived from the aforementioned SED model fits. By obtaining all properties from the same SED model, we ensure they are self-consistent, free from biases introduced when using different filters or observations to calculate different AGN or host galaxy properties, and decrease the effect of AGN variability on measured properties.

In this Section we discuss the derivation of AGN and host galaxy properties. In Section \ref{subsec: redshift} we discuss our sample's redshift distribution, and the selection of our final sample of 2585 AGN host galaxies included in the catalog. In Section \ref{subsec: AGN Lbol} we discuss the derivation of AGN bolometric luminosity from both the SED model and X-ray data, and we compare the two resulting datasets. In Section \ref{subsec: Mstar} we discuss the derivation of stellar mass from the SED model. In Section \ref{subsec: SFR} we discuss the derivation of SFR from the SED model, and in Section \ref{subsec: Nh} we discuss the derivation of column density from the SED model. All of these properties are included in the ACS-AGN catalog, as shown in Table \ref{tab: catalog}. Finally, in Section \ref{subsec: compare} we examine our derived properties in comparison with other recent work.

\subsection{Redshift ($z$)}\label{subsec: redshift}
We selected our initial sample to only be galaxies that the ACS-GC provides redshift values for, either photometric or spectroscopic. All but 11 galaxies in our final sample have photometric redshifts, and approximately 20\% of our galaxies also have spectroscopic redshifts. When a galaxy has both photometric and spectroscopic redshifts, we us the spectroscopic redshift, except in rare cases when the error on the spectroscopic redshift is large, as indicated in the ACS-GC. 

We then examine the subsample of our galaxies for which we have both photometric and spectroscopic redshifts. The ACS-GC also provides errors for their photometric redshifts; using these, and again instituting a 10\% error floor on the error values similar to the one discussed in Section \ref{subsec: SED fitting} for photometric flux errors, we find that the majority (75\%) of spectroscopic and photometric redshifts agree to within 3$\sigma$. However, there is significant deviation when examining the distributions of our lowest ($z_{\rm spec}<0.2$) and highest ($z_{\rm spec}>2.5$) redshift galaxies (median $z_{\rm spec}-z_{\rm phot}=1.0$ and 2.3, respectively). Therefore, for the rest of this paper we only study the 2585 galaxies with redshifts of $0.2<z<2.5$. The distribution of differences between spectroscopic and photometric redshift for the galaxies that fall within this range can be seen in Figure \ref{plot: reds}, left. The distribution is approximately a Guassian with enhanced tails, centered on zero, and with a standard deviation of $\sigma=0.1$

The full redshift distribution of our galaxy sample is shown in Figure \ref{plot: reds}, right. When examining our AGN subsamples, we find that our IR-selected AGNs display a flat distribution of redshifts, while the X-ray-selected AGNs display a positively skewed (longer tail towards higher redshifts) distribution and peaks at a lower redshift value than the IR-selected distribution. These distributions coincide with the previously discussed findings of \cite{Eckart2010} (see Section \ref{sec:AGN select}) that X-ray AGNs not selected in the IR tend to have lower redshifts on average. They attribute this to the fact that X-ray selection techniques are able to select low-luminosity AGNs missed by IR selection techniques.

\subsection{AGN Bolometric Luminosity ($L_{\rm AGN}$)}\label{subsec: AGN Lbol}
In order to examine SMBH growth, we need to measure AGN bolometric luminosity. Commonly, other works use the X-ray luminosity, either directly or converted to a bolometric luminosity, of the AGN when comparing SMBH growth to a host galaxy's SFR. While this attempts to avoid possible contamination added by star formation, and error added through any bolometric correction when using X-ray luminosity directly, this measurement is susceptible to AGN variability in the X-ray \citep[e.g.,][]{Hickox2014,Volonteri2015a,Volonteri2015b}. Deriving AGN bolometric luminosity by integrating an AGN SED model that has been simultaneously fit alongside a galaxy SED model, including any star formation components, allows us to avoid star formation contamination and bolometric correction error, while also being less susceptible to AGN variability due to the broadband nature of the SED.

In order to calculate the bolometric luminosity from the SED fits ($L_{\rm AGN,SED}$) we integrate the best fit rest-frame unextincted AGN SED component template for each galaxy from 0.1 $\mu$m -- 30 $\mu$m. This range dominates the integrated luminosity of the AGN and can be used as a good estimate of the AGN bolometric luminosity \citep{Assef2011}. The distribution of these bolometric luminosities are shown in Figure \ref{plot: gal props 1}, top left. Both IR and X-ray-selected AGN distributions are roughly symmetric and peak near luminosities of $10^{45}$ erg s$^{-1}$. 

For the subsample of our galaxies for which we had X-ray data, we calculate an alternate bolometric luminosity ($L_{\rm AGN,X}$) from the previously calculated $L_{2-10}$ (Section \ref*{subsec:XRAY}) using the relation found in \cite{Marconi2004}. This work improves upon earlier work by \cite{Elvis1994} and applies a larger bolometric correction for higher luminosity values. The distribution of these bolometric luminosities are shown in Figure \ref{plot: gal props 1}, top right. The IR-selected AGN subsample distribution peaks at a higher luminosity than that of the total X-ray-selected AGN distribution, indicating that AGNs selected in both the IR and X-ray have higher luminosities on average than AGNs selected in only one regime.

We also compared these two methods of calculating AGN bolometric luminosity for the 1676 (58\%) AGN host galaxies that had \textit{Chandra} data. To compare them, we examined $L_{\rm AGN,SED}$ as a function of $L_{\rm AGN,X}$ (Figure \ref{plot: gal props 1}, bottom); from this we find that the data are evenly distributed across the $L_{\rm AGN,X}=L_{\rm AGN,SED}$ line for all luminosities, and therefore the difference distribution has no dependence on AGN luminosity. We find that the distribution of log-differences in AGN bolometric luminosity between the two methods is a Gaussian centered at zero with a standard deviation of 0.6. This indicates that the integrated luminosity of the modeled AGN SED is a comparable measure to traditional X-ray bolometric correction methods on average. The spread of the distribution may be due to the assumption of a uniform photon index value when calculating $L_{\rm AGN,X}$, causing obscuration, and therefore luminosity, to be under or overestimated in specific cases.
	
This can be further investigated by examining reported hardness ratios (HR) for our galaxies. HR is a metric which roughly quantifies the X-ray spectral shape of an X-ray source, specifically the intrinsic absorption \citep{Marchesi2016}, where low values of HR represent less absorption and higher values represent more absorption. The photon index used when calculating X-ray luminosity in Section \ref{subsec:XRAY} also assumes an X-ray spectral shape and intrinsic absorption, with a value of 1.9 being an accepted value for an unabsorbed AGN \citep{Corral2011}. Therefore, our choice of 1.7 is an assumption of slight absorption in our AGN.

To investigate whether the scatter could be due to the uniform photon index assumption, we examined the differences in AGN bolometric luminosity as a function of HR for the subset of our sample which had HR values reported in the COSMOS-Legacy catalog \citep{Civano2016}, and find that there is only partial correlation between the two. Specifically, when binning by HR, we find that at low HR (HR $<-0.4$) our SED model AGN luminosity is greater when compared to the X-ray calculated AGN bolometric luminosity by approximately 0.2 dex (with a standard deviation of 0.4 dex). This follows with our assumptions, as at low HR our choice of photon index was too low, causing an overestimation of X-ray absorption. But at higher HR (HR $>-0.4$) the two calculations for AGN bolometric luminosity are in agreement (no difference but with standard deviations up to 0.6 dex). If the scatter in our two methods was purely due to the use of a single photon index, at the highest HR values we would expect the SED model AGN luminosity to be less than the X-ray calculated AGN bolometric luminosity as our chosen photon index should be underestimating X-ray absorption in these cases. Since we do not see this, this is either a selection effect in the comparison sample (we do note a lack of high HR sources, with only $7\%$ of the comparison sample having values of HR $>0.4$) or there must be further scatter intrinsic in the X-ray calculation of AGN bolometric luminosity than can be accounted for solely by the use of a single photon index. The lack of X-ray data for a significant portion of our sample and the scatter introduced from the assumption of a uniform photon index when calculating AGN luminosities from X-ray data led us to choose $L_{\rm AGN,SED}$ as the AGN luminosity metric used in our analysis for the rest of this paper.

\subsection{Stellar Mass ($M_{*}$)}\label{subsec: Mstar}
We used a color to specific luminosity relation in order to derive our galaxy stellar masses. First, we obtained the modeled $g'$ and $r'$ band galaxy magnitudes as well as the restframe luminosity in the $r$ band ($L_{r}$), not including the contribution from the AGN component, from the SED model. We then used these to calculate galaxy stellar mass using the relation between $g'-r'$ color and $M/L_{r}$ from \cite{Bell2003}: log$_{10}(M/L_{r}) = -0.499 + 1.519(g'-r')$, where the $M/L_{r}$ ratio is in solar units. 

The stellar mass distribution of our active galaxy sample is shown in Figure \ref{plot: gal props 2}, top left. We see that the X-ray-selected AGN host galaxy distribution peaks at a higher stellar mass than that of the IR-selected AGN host galaxy distribution. These findings are in line with the previously discussed findings of \cite{Azadi2017}, that X-ray-selected AGN host galaxies are biased towards higher stellar mass values, while IR-selected AGN host galaxies are biased towards moderate mass values (see Section \ref{sec:AGN select}).

\begin{figure*}
	\centering
	\includegraphics[width=0.47\textwidth]{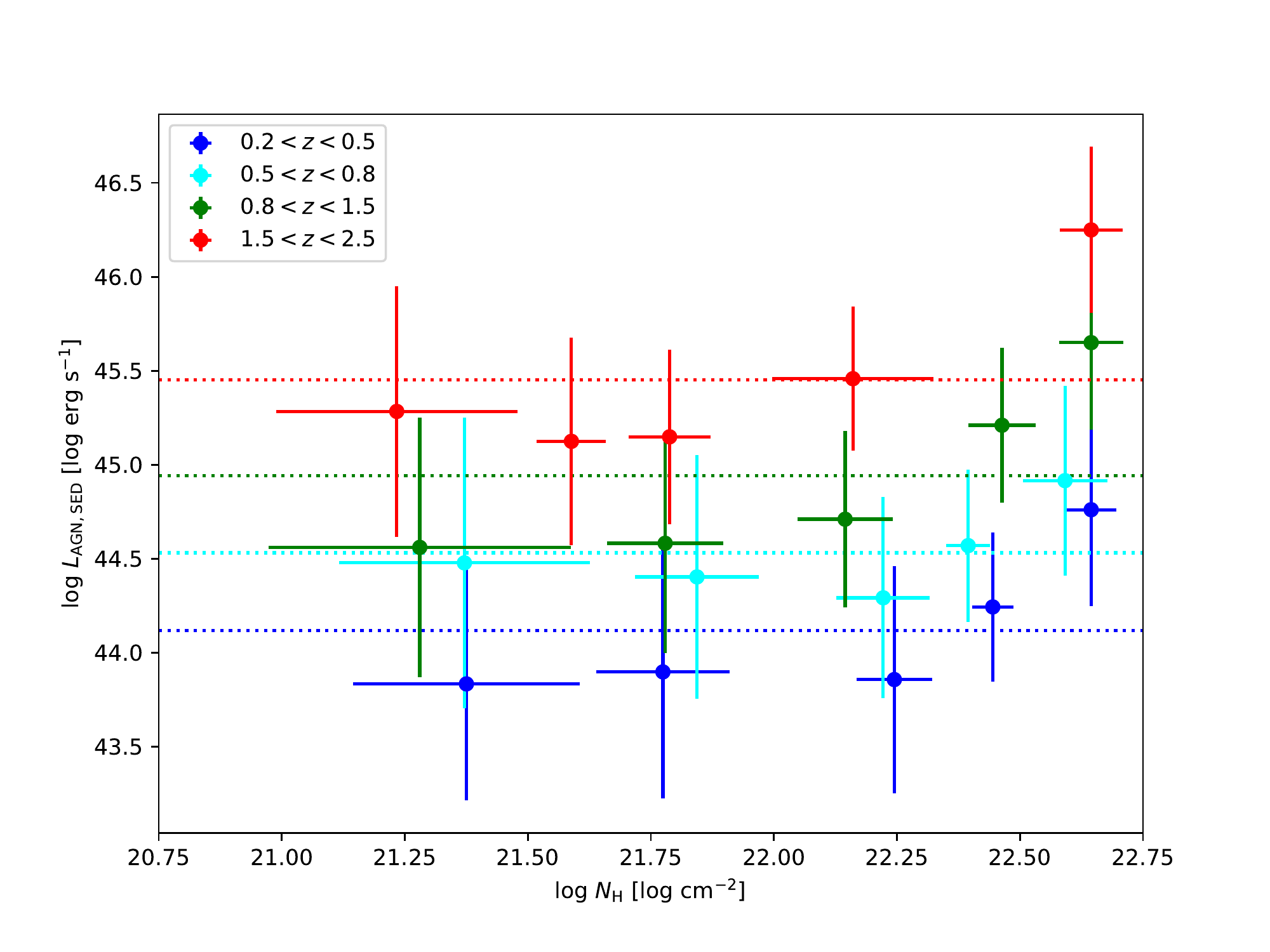}
	\includegraphics[width=0.47\textwidth]{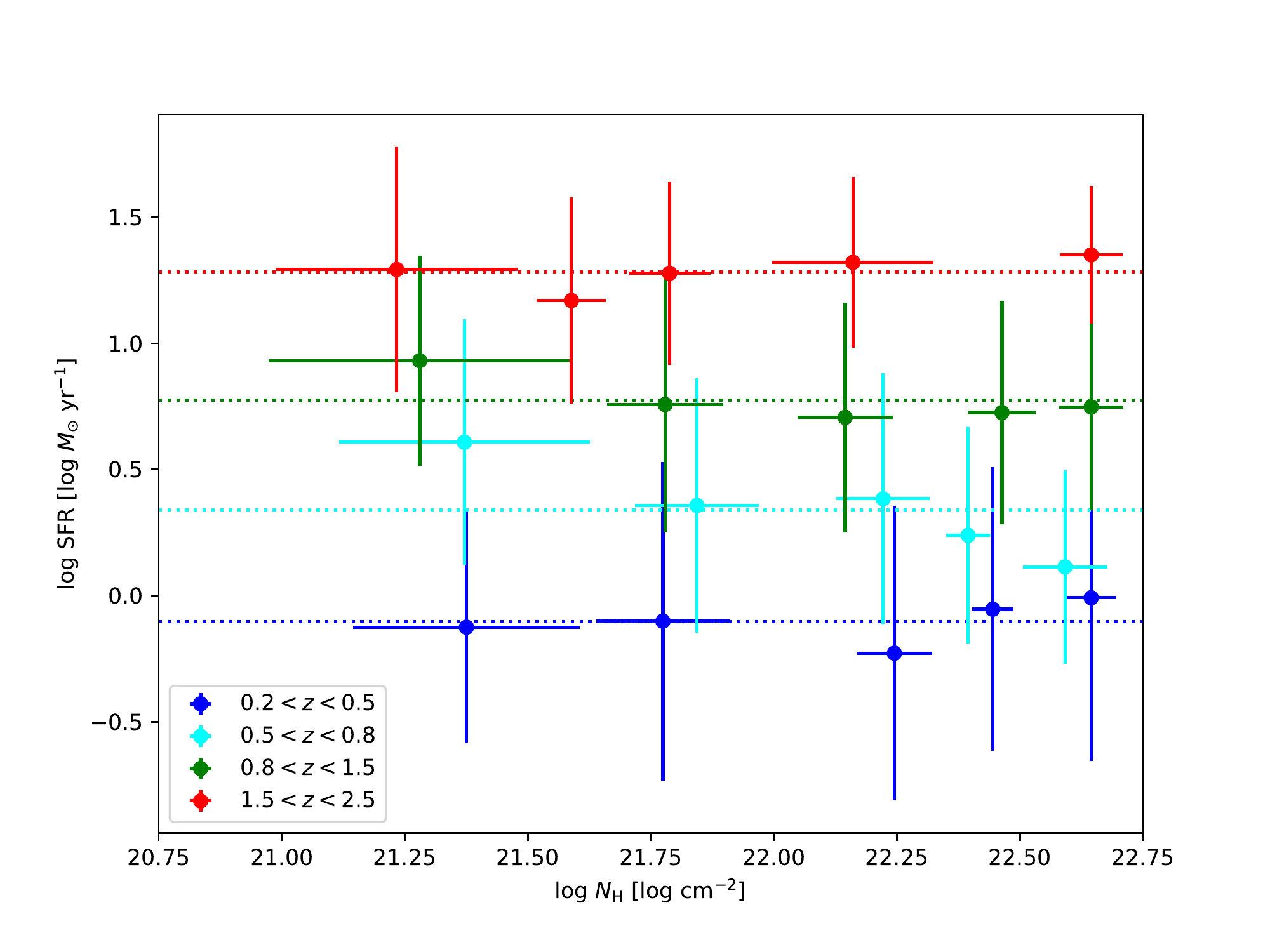}
	
	\caption{A plot of AGN bolometric luminosity as a function of nuclear column density for our active galaxy sample (left), binned by redshift (blue $=0.2<z<0.5$, cyan $=0.5<z<0.8$, green $=0.8<z<1.5$, and red $=1.5<z<2.5$) and then by nuclear column density such that each column density bin has an equal number of galaxies for a given redshift bin; the horizontal dashed lines are powerlaw slopes of zero (constant). Note that the points are consistent with zero slope, but appear to begin to show a positive correlation above column densities of $\sim$$10^{22}$ cm$^{-2}$. A plot of host galaxy star formation rate as a function of nuclear column density, binned by redshift and then nuclear column density, for our active galaxy sample (right), where the horizontal dashed lines are powerlaw slopes of zero (constant); note that the points are consistent with zero slope throughout.}
	\label{plot: gal props 3}
	\vspace{12pt}
\end{figure*}

\subsection{Star Formation Rate (SFR)}\label{subsec: SFR}
There are multiple methods used to calculate a galaxy's SFR, including emission line diagnostics, monochromatic luminosities, and integrated IR luminosity. Since we have modeled our galaxy SEDs, we opt to use a monochromatic luminosity conversion to estimate SFRs.  This approach has the added benefit of not having contamination from the AGN, since each SED model included separate AGN and galaxy components. We used the 2800 \AA$ $ monochromatic luminosity conversion, assuming a Salpeter IMF, described in \cite{Madau1998} in order to calculate our host galaxy SFRs: SFR/($M_{\odot}$ yr$^{-1}$) $= L_{2800}/(7.9\times10^{27}$ erg s$^{-1}$ Hz$^{-1}$). The distribution of these derived values is shown in Figure \ref{plot: gal props 2}, top right. Comparing the two AGN selection methods, it appears that galaxies hosting IR-selected AGNs have slightly higher SFR values on average than galaxies hosting X-ray-selected AGNs (we discuss this further in Section \ref*{subsec: AGN below SFR MS}): IR-selected AGN host galaxy median SFR $=8.5\pm0.6$ $M_{\odot}$ yr$^{-1}$ and X-ray-selected AGN host galaxy median SFR $=5.9\pm0.3$ $M_{\odot}$ yr$^{-1}$.

By combining our stellar mass and SFR measurements, we also examine specific star formation rates (sSFR $\equiv$ SFR/$M_{*}$). Figure \ref{plot: gal props 2}, bottom left, shows that the two selection methods are significantly diverged, with the IR-selected AGN host galaxy distribution having a strong negative skew and peaking more than an order of magnitude higher in sSFR values than the X-ray-selected AGN host galaxy distribution.

\subsection{Nuclear Column Density ($N_{\rm H}$)}\label{subsec: Nh}
One of the parameters fit by \texttt{LRT} is an extinction value, $E_{B-V}$, that is applied to the AGN template. The $E_{B-V}$ extinction parameter can be converted to column density via a conversion derived from two studies \citep{Maiolino2001,Burtscher2016} that found a ratio between the two for approximately 40 AGN host galaxies. By computing a weighted average of all galaxies presented in the two works, we find a conversion factor of $E_{B-V}/N_{\rm H}=1.80\pm0.15\times10^{-23}$ cm$^{2}$. After applying this conversion, we find the distribution shown in Figure \ref{plot: gal props 2}, bottom right. We find that our IR-selected AGN distribution is significantly bimodal, with our X-ray-selected AGN distribution also showing a secondary peak at the same location as that of the IR-selected AGN distribution. This bimodal nature of column density values for a population of AGNs was observed previously by \cite{Civano2016}, which used the hardness ratio in place of column density. \cite{Civano2016} interpreted this bimodal feature as the result of the galaxy population containing both obscured and unobscured AGNs.

Further, we analyze trends in the comparison of our column density values as a function of known HR values from \cite{Civano2016}. As discussed in Section \ref{subsec: AGN Lbol}, low HR values correspond with less intrinsic absorption (low column densities) and high HR values correspond to more intrinsic absorption (high column densities). Therefore as HR values increase, we should expect to see column densities increase as well. When examining our AGN host galaxies, we find that while our column densities do increase as HR becomes more positive, there is considerable scatter, with standard deviations in column density of approximately 0.5 dex across all HR bins.

We also examine the relationship between nuclear column density and AGN luminosity (Figure \ref{plot: gal props 3}, left), as well as host galaxy SFR (Figure \ref{plot: gal props 3}, right). First, we find both the $L_{\rm AGN,SED}-N_{\rm H}$ and SFR -- $N_{\rm H}$ relations are consistent with a powerlaw slope of zero (constant) for all redshift bins (none of the eight powerlaw fits resulted in a p-value $<$ 0.01, and only one with a p-value $<$ 0.05). However, we do note that at higher nuclear column densities ($N_{\rm H}\gtrsim10^{22}$ cm$^{-2}$), the AGN bolometric luminosity seems to increase with nuclear column density. This behavior seems to be apparent at all redshifts, but is not significant in any (none of the eight powerlaw fits using only data points with $N_{\rm H}\gtrsim10^{22}$ cm$^{-2}$ resulted in a p-value $<$ 0.01, and only one with a p-value $<$ 0.05). This may indicate that nuclear column density is not a critical tracer of AGN fueling until reaching higher values or could be a result of the flux-limited nature of our sample; possible future work could explore whether this trend is significant at higher column densities and into the Compton thick regime ($N_{\rm H}\geq10^{24}$ cm$^{-2}$) for a sample which is not flux-limited. 

The lack of correlation between nuclear column density and host galaxy star formation has been observed before \citep[e.g.,][]{Rosario2012}, and may point to a difference in fueling processes or timescales between SMBH growth and host galaxy star formation. This is surprising however given that models by \cite{Sanders1988} predict that most SMBH growth occurs when the AGN is surrounded by a dense, dusty, obscuring envelope, and models by \cite{Somerville2008} predict that AGN obscuration should trend with a galaxy's global SFR.

\subsection{Comparison to Other Work}\label{subsec: compare}
The variety and depth of observations in the COSMOS field has enabled a significant number of works and catalogs to be built from them. The catalog produced by \cite{Jin2018} (hereafter J18) is one such work; galaxy properties within this catalog, including SFR and $M_{*}$, are derived from SED fits to ``super-deblended" photometric observations ranging in wavelength from the far-IR (FIR) to the submillimeter (sub-mm). Matching AGN host galaxies from our catalog to that of J18 using a search cone of radius 0\farcs25, and requiring a difference in redshift of $\Delta z<0.1$,  results in 1197 galaxies with which we can compare derived SFR and $M_{*}$ values.

When comparing our galaxy properties to those derived by J18, we find that our log($M_{*}$) values are systematically higher than those of J18, with a median difference of 0.25 dex, and that our log(SFR) values are systematically lower, with a median difference of -0.71 dex. These disagreements are due to differences in methods between this work and J18. The first significant difference is the bands used to create the SEDs in each work, with J18 using FIR to sub-mm bands and this work using mid-UV to mid-IR bands. This results in minimal overlap in the observational wavelength regime used to build the underlying SEDs and derive galaxy properties. Secondly, J18 derive their SFR via a two-step process. First an initial SED fit is used to calculate an initial SFR, which is then used to choose whether a starburst or main-sequence galaxy type stellar component is used in the final SED fit. This final SED fit, based on the preliminary SFR measurement, is then used to determine the final SFR. This approach limits the range of stellar component fits while the method described in Section \ref{subsec: SED fitting} of this work allows for a large range of stellar component SEDs by using a linear combination of three diverse galaxy templates for all SED fits. This could explain the lower SFR values found in this work when compared to the values found in J18.

Further, our SED fitting method and derivation of galaxy properties has been tested by \cite{Barrows2017b}. They examine four SDSS galaxies and compare SFR and $M_{*}$ values derived using the same method as this work to measurements made with SDSS optical fiber spectra and included in the MPA-JHU catalog \citep{Kauffmann2003,Brinchmann2004}. \cite{Barrows2017b} find their derived values of SFR and $M_{*}$ to be consistent within their uncertainties to those of the MPA-JHU catalog and they find no systematic offset between the two (see \cite{Barrows2017b} for a more in-depth discussion). 

Therefore, we find it difficult to make a direct comparison to J18 because of significant differences in methods between our two works. We believe our approach to be more appropriate due to our use of a more diverse wavelength range and a more diverse set of SED stellar components when creating the SED models from which we calculate SFR and $M_{*}$ values. Our approach is also consistent with SFR and $M_{*}$ values in SDSS as shown in \cite{Barrows2017b}.

\begin{deluxetable*}{lllll}[t!]
	\tablewidth{0pt}
	\tablecolumns{3}
	\tablecaption{Data Fields in the ACS-AGN Catalog\label{tab: catalog}}
	\tablehead{
		\colhead{No. $\;\;\;$}&
		\colhead{Field $\;\;\;\;\;\;\;\;\;\;\;\;\;\;\;\;\;\;\;\;$} &
		\colhead{Note}
	}
	\startdata
	1 & ID & catalog specific unique identifier \\
	2 & RA & right ascension [J2000, decimal degrees] \\
	3 & DEC & declination [J2000, decimal degrees] \\
	4 & Z & redshift used \\
	5 & SPECZ & spectroscopic redshift \\
	6 & PHOTOZ & photometric redshift \\
	7 & Spitzer{\_}AGN & if AGN was selected in \textit{Spitzer} [Boolean] \\
	8 & Chandra{\_}AGN & if AGN was selected in \textit{Chandra} [Boolean] \\
	9 & L{\_}bol{\_}sed{\_}md & AGN bolometric luminosity calculated from SED, median [erg s$^{-1}$] \\
	10 & L{\_}bol{\_}sed{\_}lo & AGN bolometric luminosity calculated from SED, lower bound [erg s$^{-1}$] \\
	11 & L{\_}bol{\_}sed{\_}hi & AGN bolometric luminosity calculated from SED, upper bound [erg s$^{-1}$] \\
	12 & L{\_}x{\_}md & 2 -- 10 keV restframe luminosity, median [erg s$^{-1}$] \\
	13 & L{\_}x{\_}lo & 2 -- 10 keV restframe luminosity, lower bound [erg s$^{-1}$] \\
	14 & L{\_}x{\_}hi & 2 -- 10 keV restframe luminosity, upper bound [erg s$^{-1}$] \\
	15 & L{\_}bol{\_}x{\_}md & AGN bolometric luminosity calculated from X-ray, median [erg s$^{-1}$] \\
	16 & L{\_}bol{\_}x{\_}lo & AGN bolometric luminosity calculated from X-ray, lower bound [erg s$^{-1}$] \\
	17 & L{\_}bol{\_}x{\_}hi & AGN bolometric luminosity calculated from X-ray, upper bound [erg s$^{-1}$] \\
	18 & M{\_}star{\_}md & galaxy stellar mass, median [$M_{\odot}$] \\
	19 & M{\_}star{\_}lo & galaxy stellar mass, lower bound [$M_{\odot}$] \\
	20 & M{\_}star{\_}hi & galaxy stellar mass, upper bound [$M_{\odot}$] \\
	21 & SFR{\_}md & star formation rate, median [$M_{\odot}$ yr$^{-1}$] \\
	22 & SFR{\_}lo & star formation rate, lower bound [$M_{\odot}$ yr$^{-1}$] \\
	23 & SFR{\_}hi & star formation rate, upper bound [$M_{\odot}$ yr$^{-1}$] \\
	24 & Nh{\_}md & nuclear column density, median [cm$^{-2}$] \\
	25 & Nh{\_}lo & nuclear column density, lower bound [cm$^{-2}$] \\
	26 & Nh{\_}hi & nuclear column density, upper bound [cm$^{-2}$] \\
	27 & SFR{\_}norm{\_}md & normalized star formation rate, median \\
	28 & SFR{\_}norm{\_}lo & normalized star formation rate, lower bound \\
	29 & SFR{\_}norm{\_}hi & normalized star formation rate, upper bound
	\enddata	
	\tablecomments{Field numbers 2 -- 6 are taken from the ACS-GC catalog \citep{Griffith2012}. AGN selection and derivations of AGN and host galaxy properties are described throughout this paper. Lower bound and upper bound are defined as the 16th and 84th percentiles of the distribution, respectively. The ACS-AGN Catalog is available in its entirety in fits format from the original publisher; NULL values (fields with no data) are represented as -999 in the ACS-AGN Catalog. \vspace{-12pt}}
\end{deluxetable*}

\section{Results}\label{sec: results}
In this Section we discuss the primary results from statistical analysis of the properties of our AGNs and their host galaxies. In Section \ref{subsec: AGN below SFR MS} we examine the location of our AGN host galaxies with respect to the  star-forming main sequence, and in Section \ref{subsec: SFR Lbol link} we examine an SFR -- AGN luminosity relation, accounting for redshift. 

\begin{figure*}
	\centering
	\includegraphics[width=0.40\textwidth]{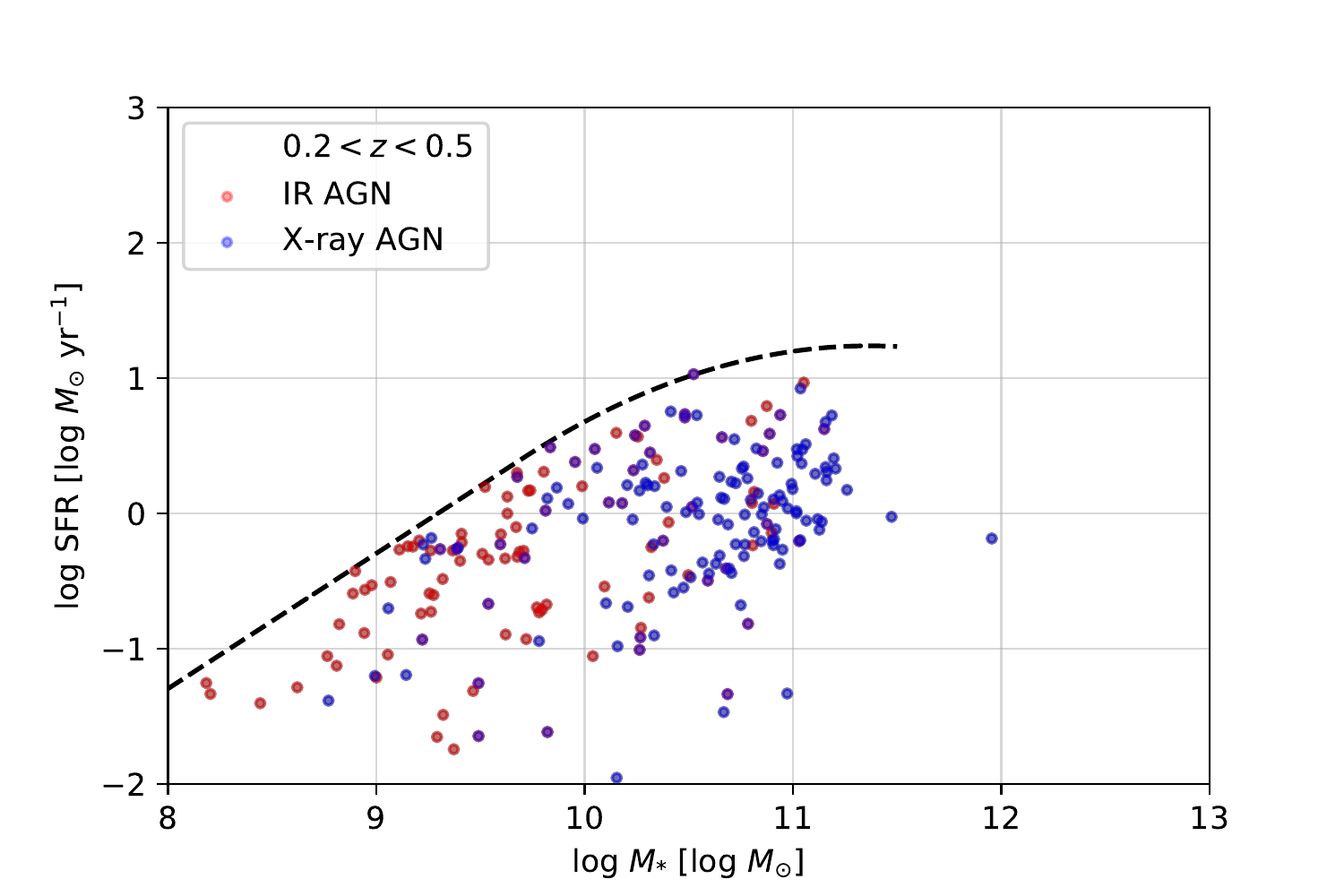}
	\includegraphics[width=0.40\textwidth]{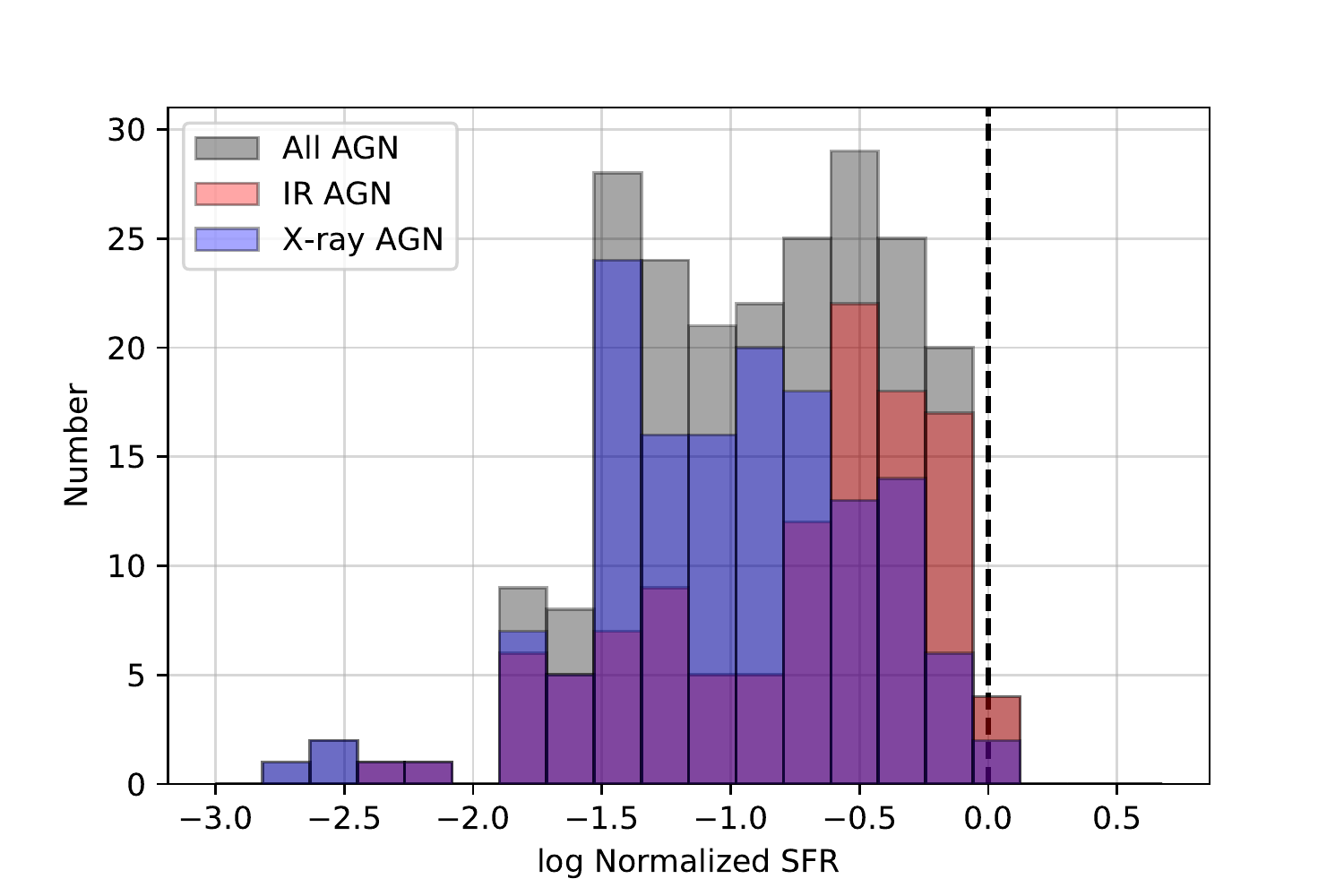}
	\includegraphics[width=0.40\textwidth]{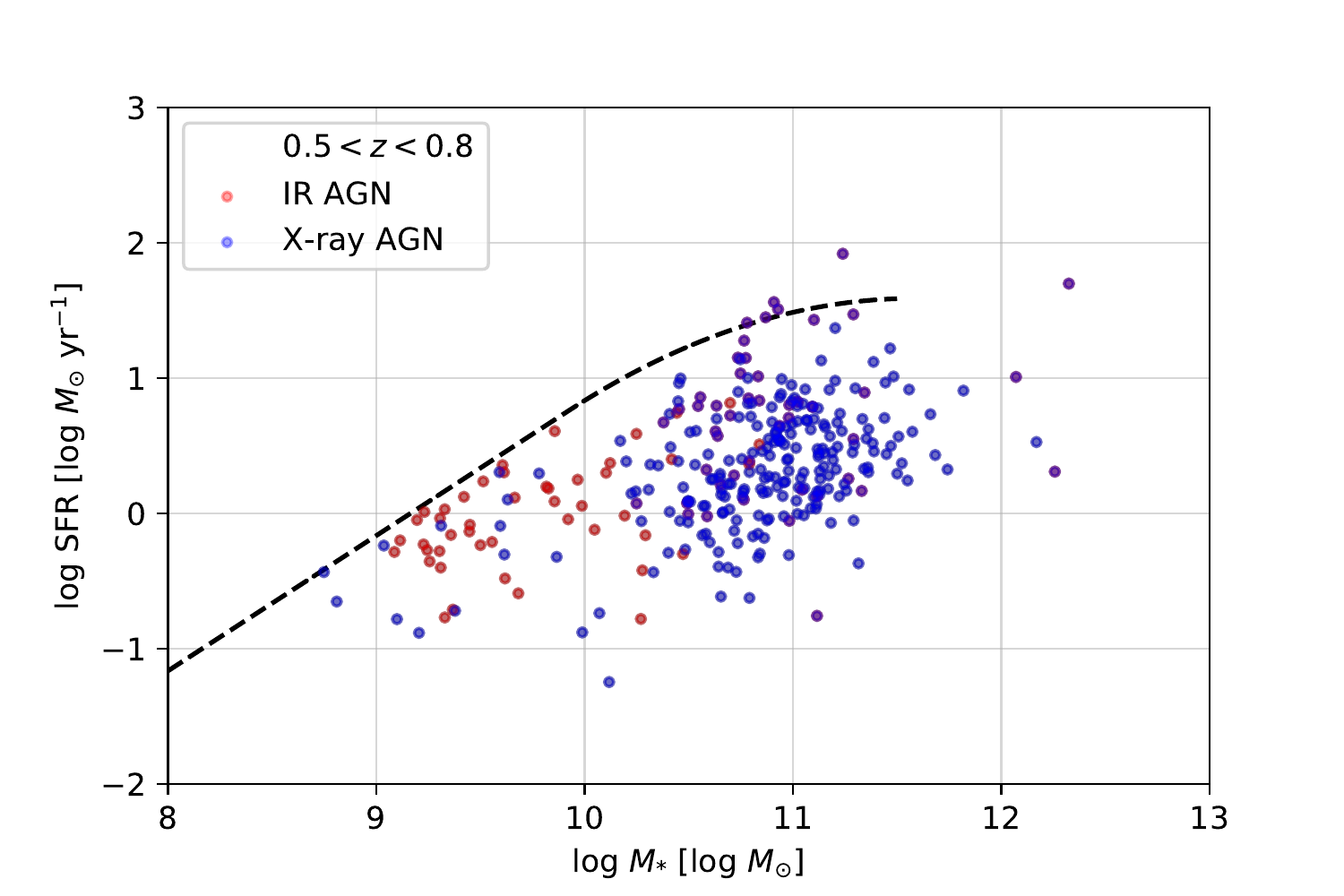}
	\includegraphics[width=0.40\textwidth]{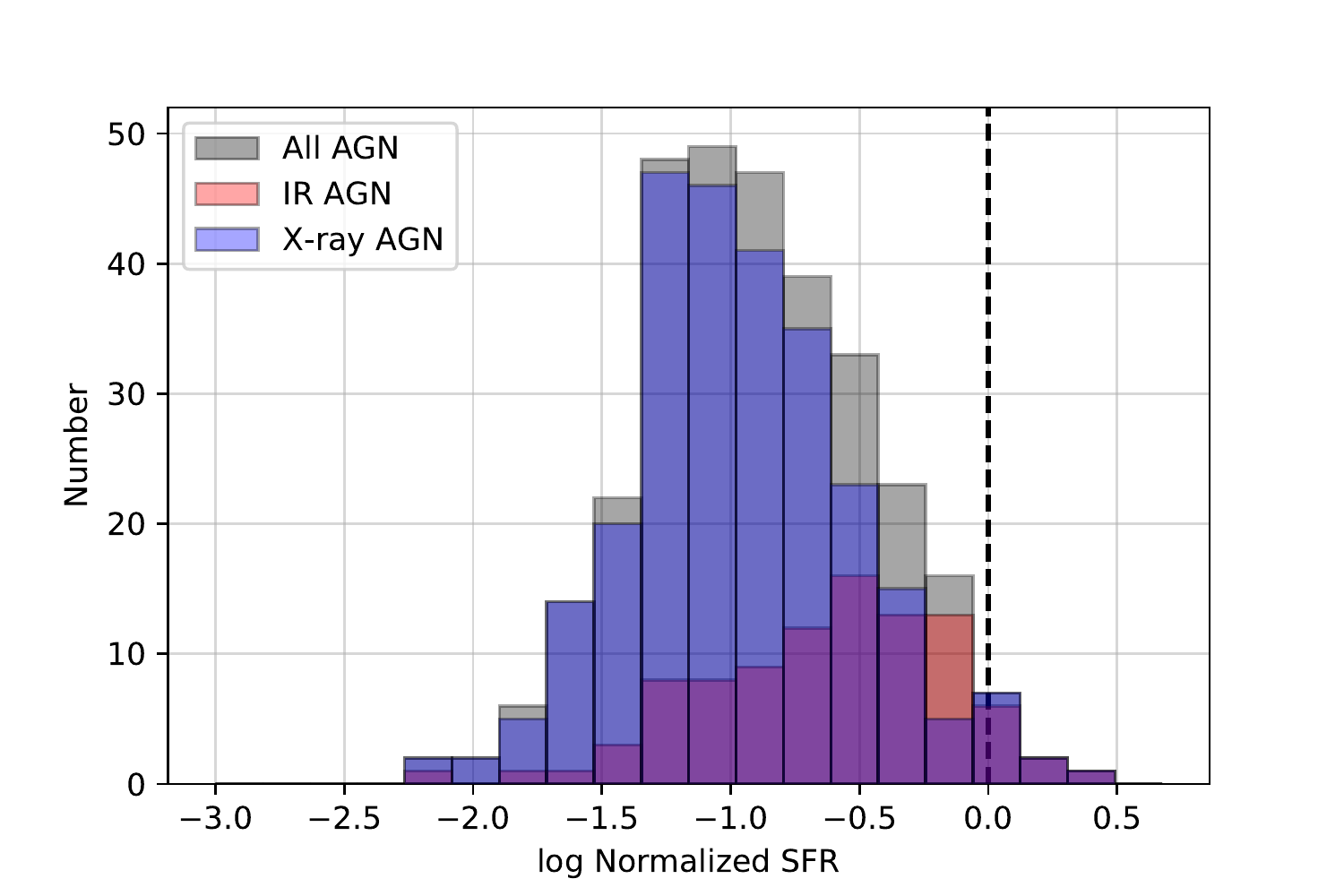}
	\includegraphics[width=0.40\textwidth]{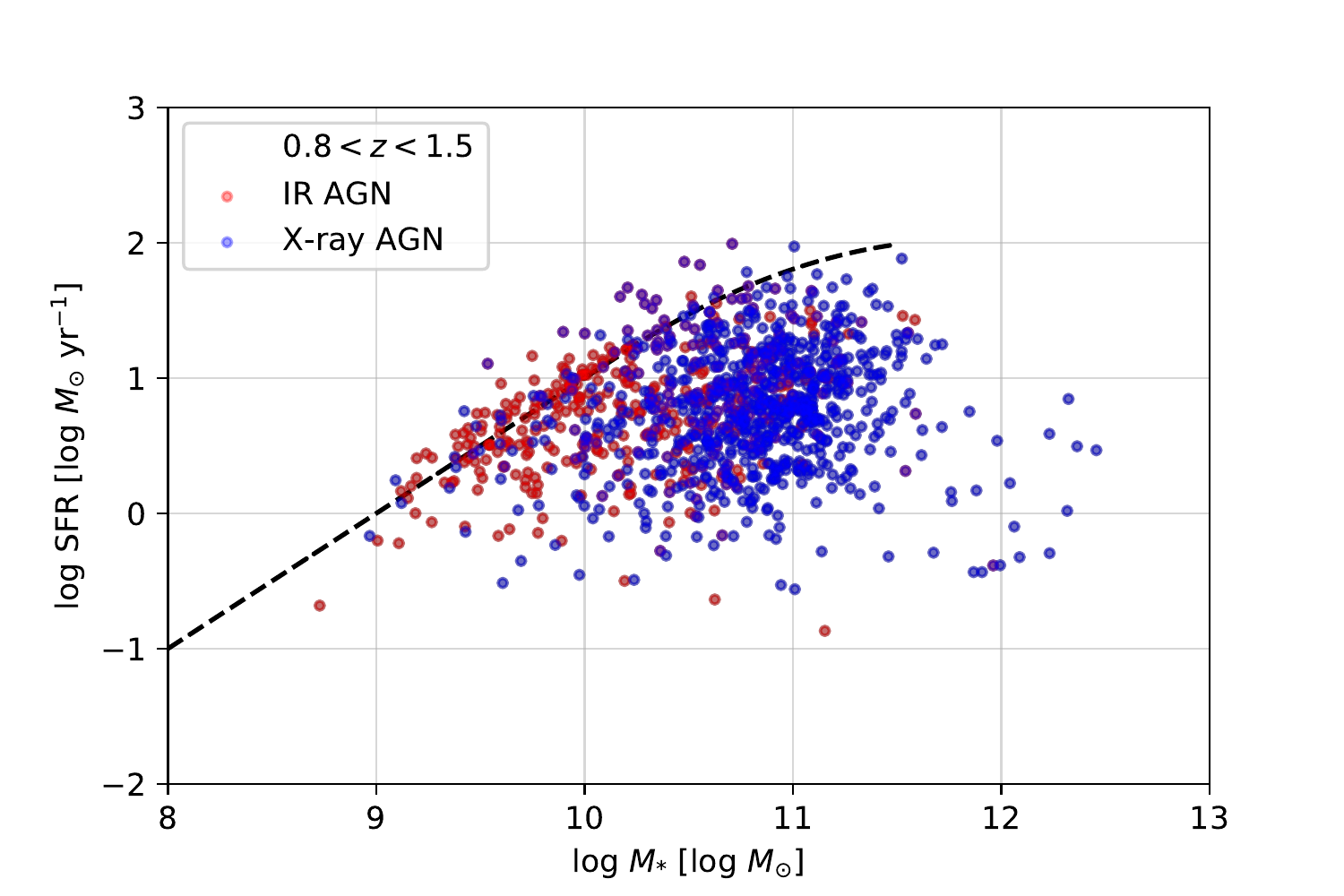}
	\includegraphics[width=0.40\textwidth]{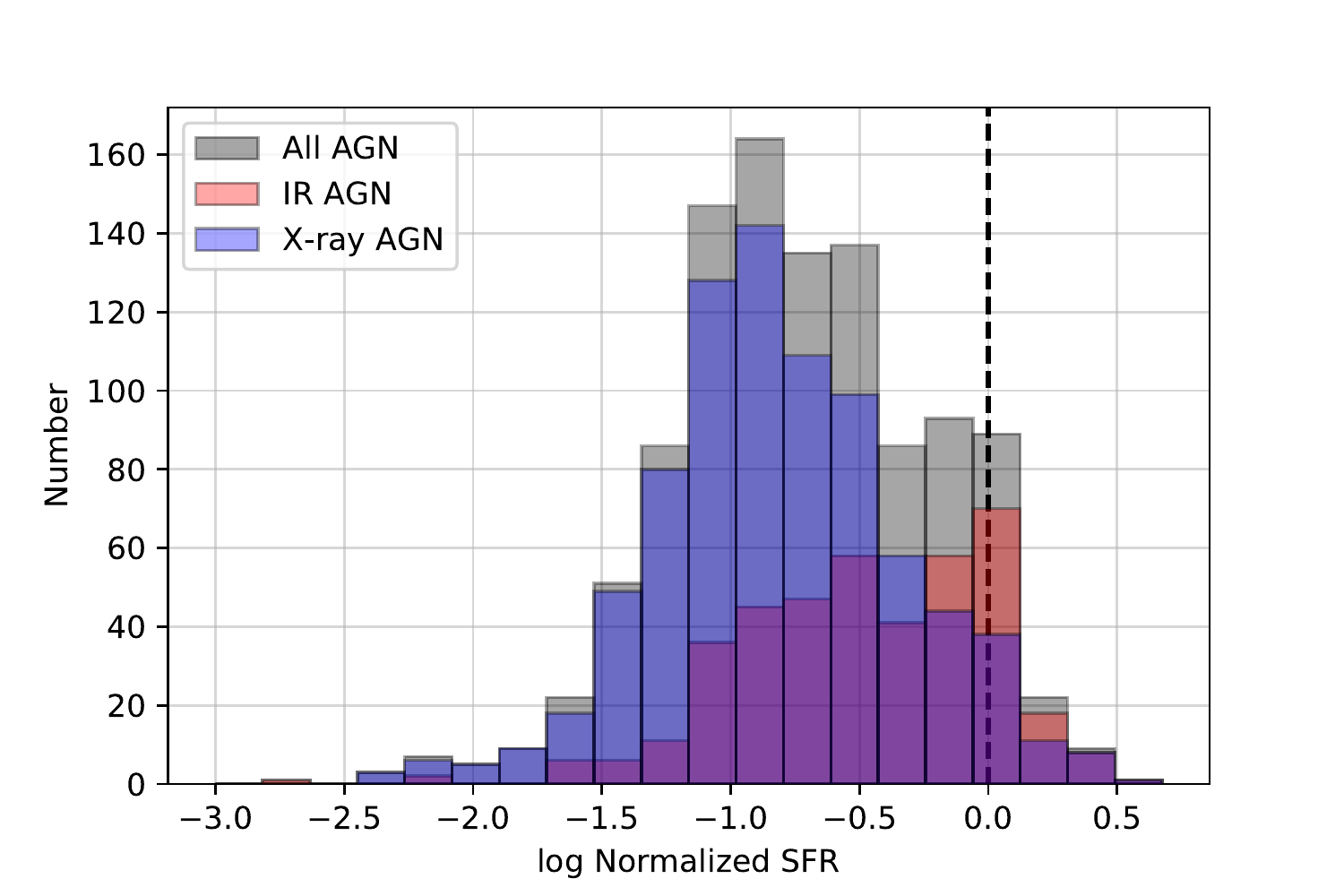}
	\includegraphics[width=0.40\textwidth]{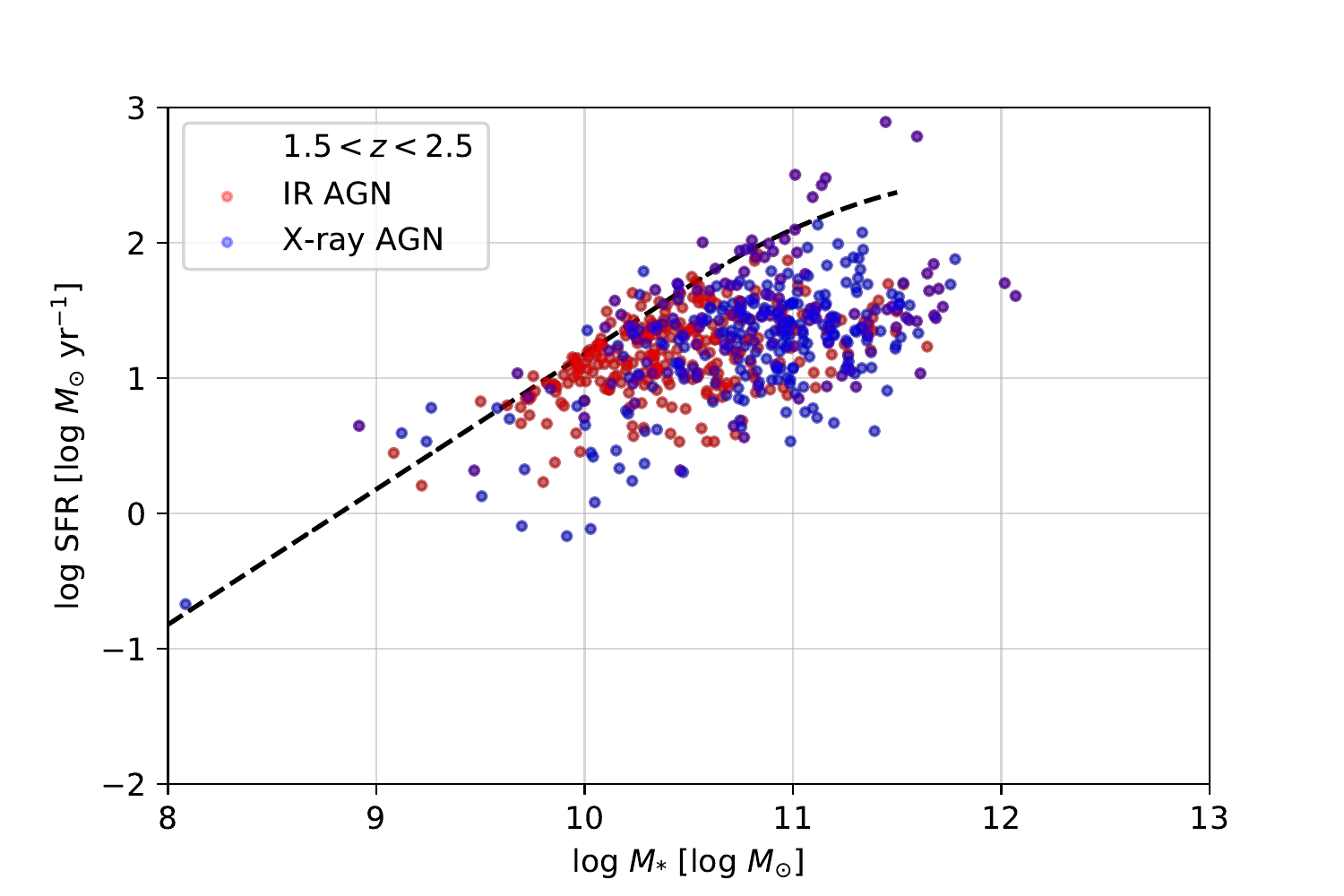}
	\includegraphics[width=0.40\textwidth]{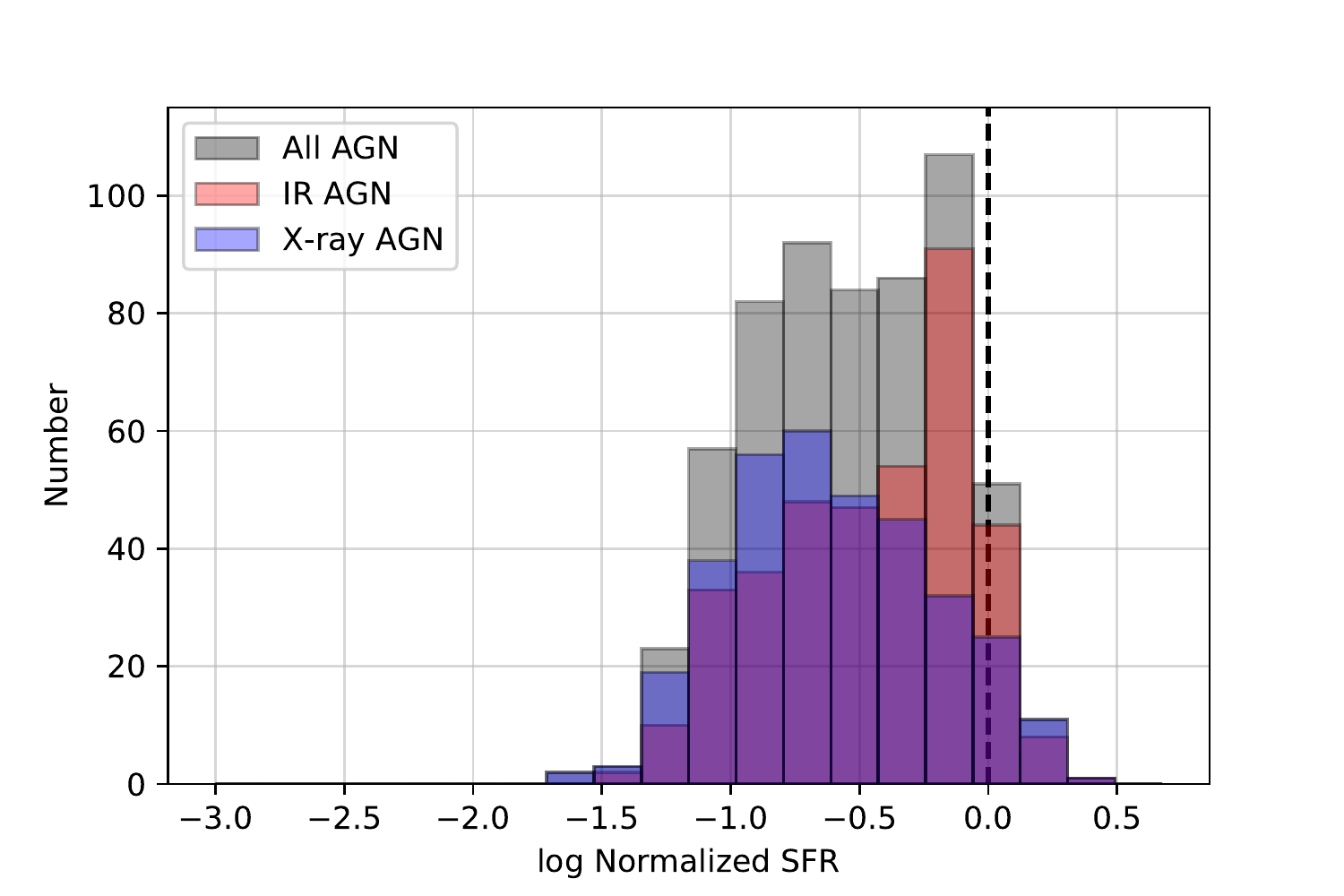}
	
	\caption{SFR as a function of stellar mass (left column) and histograms of normalized SFR $\equiv$ SFR/SFR$_{\rm MS}$ (right column) for our AGN host galaxies; each row depicts a redshift bin --- from top to bottom they are: $0.2<z<0.5$, $0.5<z<0.8$, $0.8<z<1.5$, and $1.5<z<2.5$. The dashed lines on the left plots are the star-forming main sequence from \cite{Schreiber2015} for the median redshift value in each bin, while the dashed lines on the right plots are the SFR $=$ SFR$_{\rm MS}$ line. Note that the majority of the active galaxies lie below the median star-forming main sequence for their redshift bin. This is further shown in the right plots, which show that very few of our galaxies have SFRs at or above their comparative star-forming main sequence values. We see that IR-selected AGN host galaxies have higher normalized SFR values on average than our X-ray-selected AGN host galaxies, with the two distributions' peaks typically separated by 0.5 -- 1.0 dex. Further, we find a gradual shift towards higher normalized SFRs as the galaxies increase in redshift.}
	\label{plot: results1}
	\vspace{12pt}
\end{figure*}

\subsection{Galaxies Hosting AGNs Have Lower SFRs}\label{subsec: AGN below SFR MS}
Previous studies have observed and reported on relations between a galaxy's redshift, stellar mass, and its star formation rate \citep[e.g.,][]{Noeske2007,Whitaker2012,Schreiber2015}. This relation between the star formation rate of a galaxy and its stellar mass is typically called the star-forming main sequence. This metric is a useful tool in conjunction with observed galaxy SFR in order to determine whether a given galaxy has a normal, above, or below average SFR given its redshift and mass. These relations are empirical and have been iteratively improved over the last decade \citep[e.g.][]{Whitaker2012,Schreiber2015}. We use the \cite{Schreiber2015} relation to determine the main sequence star formation rates (SFR$_{\rm MS}$) for our AGN host galaxies, defined as: log$_{10}($SFR$_{\rm MS}[M_{\odot}$ yr$^{-1}])=m-m_{0}+a_{0}r-a_{1}[max(0,m-m_{1}-a_{2}r)]^{2}$, where $r\equiv $ log$_{10}(1+z)$, $m\equiv $ log$_{10}(M_{*}/10^{9} M_{\odot})$, $m_{0}=0.5\pm0.07$, $a_{0}=1.5\pm0.15$, $a_{1}=0.3\pm0.08$, $m_{1}=0.36\pm0.3$, and $a_{2}=2.5\pm0.6$. 

We use this relation to test whether a general population of star-forming galaxies with SFRs calculated using the methods presented in this paper lies on the star-forming main sequence defined by \cite{Schreiber2015}. To do this, we randomly selected, created SEDs, and extracted galaxy properties for 2245 star-forming ACS-GC galaxies in the COSMOS field, where we use the \cite{Schreiber2015} definition of star-forming galaxies. These galaxies were randomly selected spatially from the same parent sample as our AGN sample, and have similar distributions for all measured parameters. In order to examine an individual galaxy's SFR in relation to the star-forming main sequence value, we also calculated a normalized SFR (normalized SFR $\equiv$ SFR/SFR$_{\rm MS}$). We find that these star-forming galaxies do lie predominantly on the star formation main sequence, with median (standard deviation) log normalized SFR values of $-$0.23 (0.29), $-$0.30 (0.34), $-$0.14 (0.39), and $-$0.11 (0.27) for redshift bins of $0.2<z<0.5$, $0.5<z<0.8$, $0.8<z<1.5$, and $1.5<z<2.5$, respectively.

In contrast we find that our AGN host galaxies lie, on average, below the star-forming main sequence, as can be seen in Figure \ref{plot: results1}. Specifically we find that our AGN host galaxies lie below the star-forming main sequence with median (standard deviation) log normalized SFR values of: $-$0.54 (0.55), $-$0.57 (0.48), $-$0.47 (0.48), and $-$0.39 (0.38) for IR-selected AGN host galaxies; $-$0.98 (0.52), $-$1.01 (0.44), $-$0.85 (0.47), and $-$0.62 (0.40) for X-ray-selected AGN host galaxies; and $-$0.84 (0.54), $-$0.95 (0.45), $-$0.75 (0.50), and $-$0.52 (0.39) for the total AGN host galaxy sample, for redshift bins of $0.2<z<0.5$, $0.5<z<0.8$, $0.8<z<1.5$, and $1.5<z<2.5$, respectively. 

If we account for the systematic offset of the matched inactive star-forming galaxies below the main sequence as discussed above, we find that our AGN host galaxies lie below the star-forming main sequence with median (standard deviation) log normalized SFR values of: $-$0.31 (0.62), $-$0.27 (0.59), $-$0.33 (0.62), and $-$0.28 (0.47) for IR-selected AGN host galaxies; $-$0.75 (0.60), $-$0.71 (0.56), $-$0.71 (0.61), and $-$0.51 (0.48) for X-ray-selected AGN host galaxies; and $-$0.61 (0.61), $-$0.65 (0.56), $-$0.61 (0.63), and $-$0.41 (0.47) for the total AGN host galaxy sample, for the respective redshift bins.

Examining the normalized SFRs of our galaxies seen in Figure \ref{plot: results1}, we find a similar but distinct trend to that of sSFR (see Sections \ref{subsec: Mstar} and \ref{subsec: SFR}). Because our X-ray-selected AGN host galaxies have higher masses and similar to lower SFRs in comparison to IR-selected AGN host galaxies, we find a compound effect when examining normalized SFR. While both AGN subpopulations, on average, lie below their comparative star-forming main sequence values, the IR-selected AGN host galaxies tend to be closer to the main sequence --- with median normalized SFR values lying within one standard deviation of the main sequence for all redshift bins --- than the X-ray-selected AGN host galaxies. We find this for all redshift bins, with higher redshift bins producing normalized SFRs that are closer to 1 and less discrepancy between the two AGN subpopulations.

Recent work by \cite{Bernhard2019} finds that galaxies hosting higher luminosity AGNs lie closer to the star formation main sequence than galaxies hosting lower luminosity AGNs; however our IR and X-ray-selected AGN host galaxies have similar AGN luminosities (see Figure \ref{plot: gal props 1}, top left), ruling this out as a possible explanation. Instead, the differences between these two selection techniques can be attributed to the selection biases described in \cite{Azadi2017} and discussed in Section \ref{subsec:Comparison}. Specifically, that IR selection techniques favor galaxies that are moderate mass, dusty, and have higher SFRs, while X-ray selection techniques favor galaxies that are high mass, less dusty, and quiescent. These selection biases cause IR-selected AGN host galaxies to shift up (higher SFR) on the galaxy mass -- SFR plane, while X-ray-selected AGN host galaxies are shifted right (higher mass) and down (quiescent). This separates the two populations, with IR-selected AGN host galaxies lying closer to the star formation main sequence than X-ray-selected AGN host galaxies.

Previous work has examined the location of AGN host galaxies in relation to the star-forming main sequence. Work by \cite{Stanley2017} find that on, on average, galaxies hosting AGNs lie along the star-forming main sequence, while work by \cite{Shimizu2015} and \cite{Mullaney2015} find that AGN host galaxies tend to lie below the star-forming main sequence, i.e., have less star formation than other galaxies of similar mass. 

This disagreement may stem from the wavelength regimes used to select the AGNs and measure the SFRs. Specifically, \cite{Stanley2017} select AGNs using visible observations from SDSS, and find that their AGN sample lies on or near the star-forming main sequence. In contrast to \cite{Stanley2017}, studies by \cite{Shimizu2015} and \cite{Mullaney2015} select their AGNs using X-ray observations from \textit{Swift}/BAT and \textit{Chandra}, respectively. Their findings using X-ray selection coincides with ours, that X-ray selected AGN host galaxies lie significantly below the star-forming main sequence on average.

However, it is also possible that the disagreements in the findings of previous works arise from differing derivation techniques for both stellar mass and SFR. For example, \cite{Stanley2017} uses SED models to find SFRs, and then uses emission line derived SMBH mass values in order to then extract galaxy stellar mass values using a $M_{BH}$ -- $M_{*}$ relation. Our work avoids any discrepancies introduced in using multiple observation methods to determine galaxy properties by deriving all of them using self-consistent SED fitting techniques that reproduce the \cite{Schreiber2015} star-forming main sequence when examining a general population of star-forming galaxies.

Our findings that AGN host galaxies are, on average, offset below the star-forming main sequence seem to indicate that there is a mechanism that precludes the maximal or even average formation rate of stars in AGN dominated systems. This may provide some evidence for the presence of negative feedback on star formation in these systems.

\begin{figure*}
	\centering
	\includegraphics[width=0.47\textwidth]{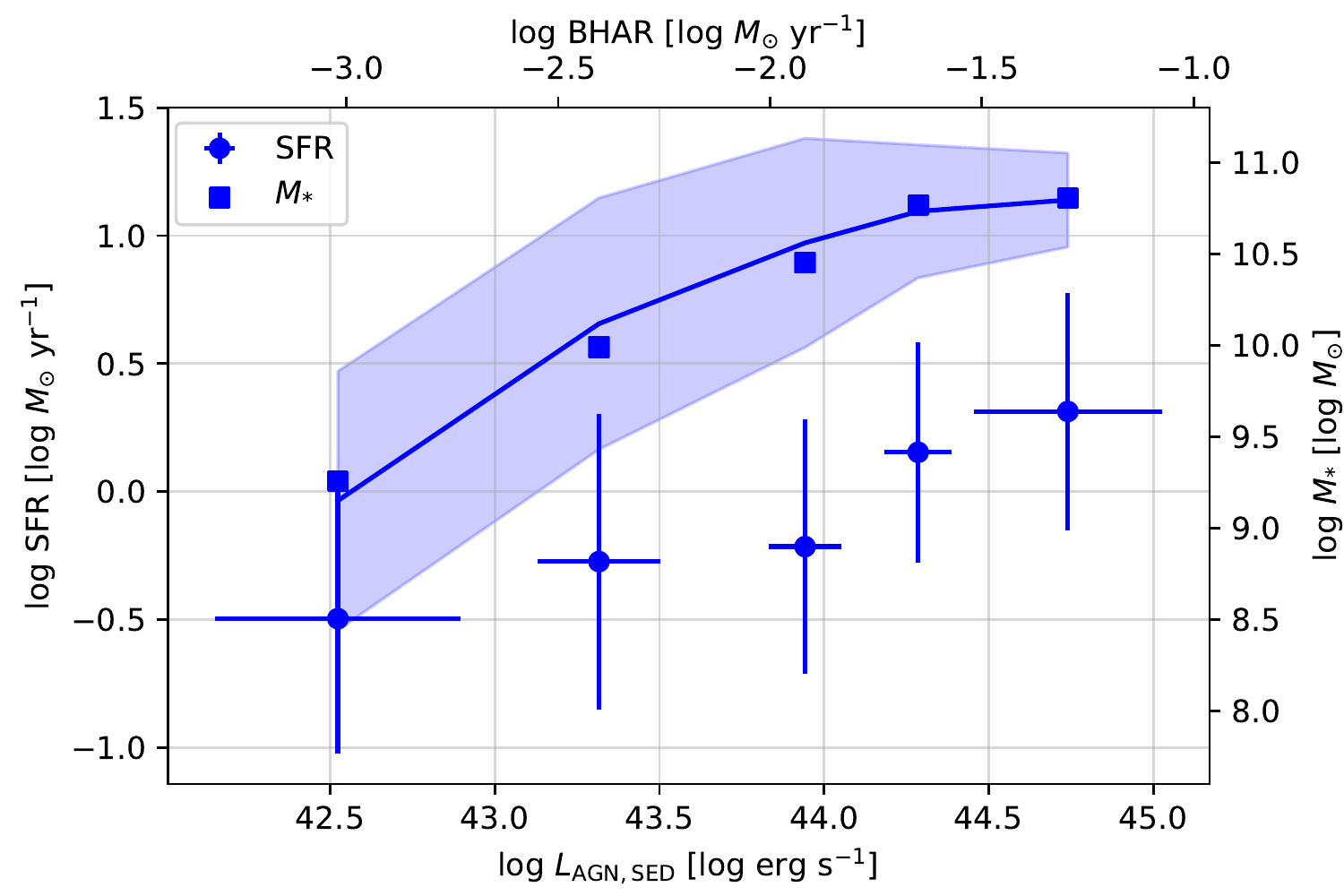} \vspace{12pt}
	\includegraphics[width=0.47\textwidth]{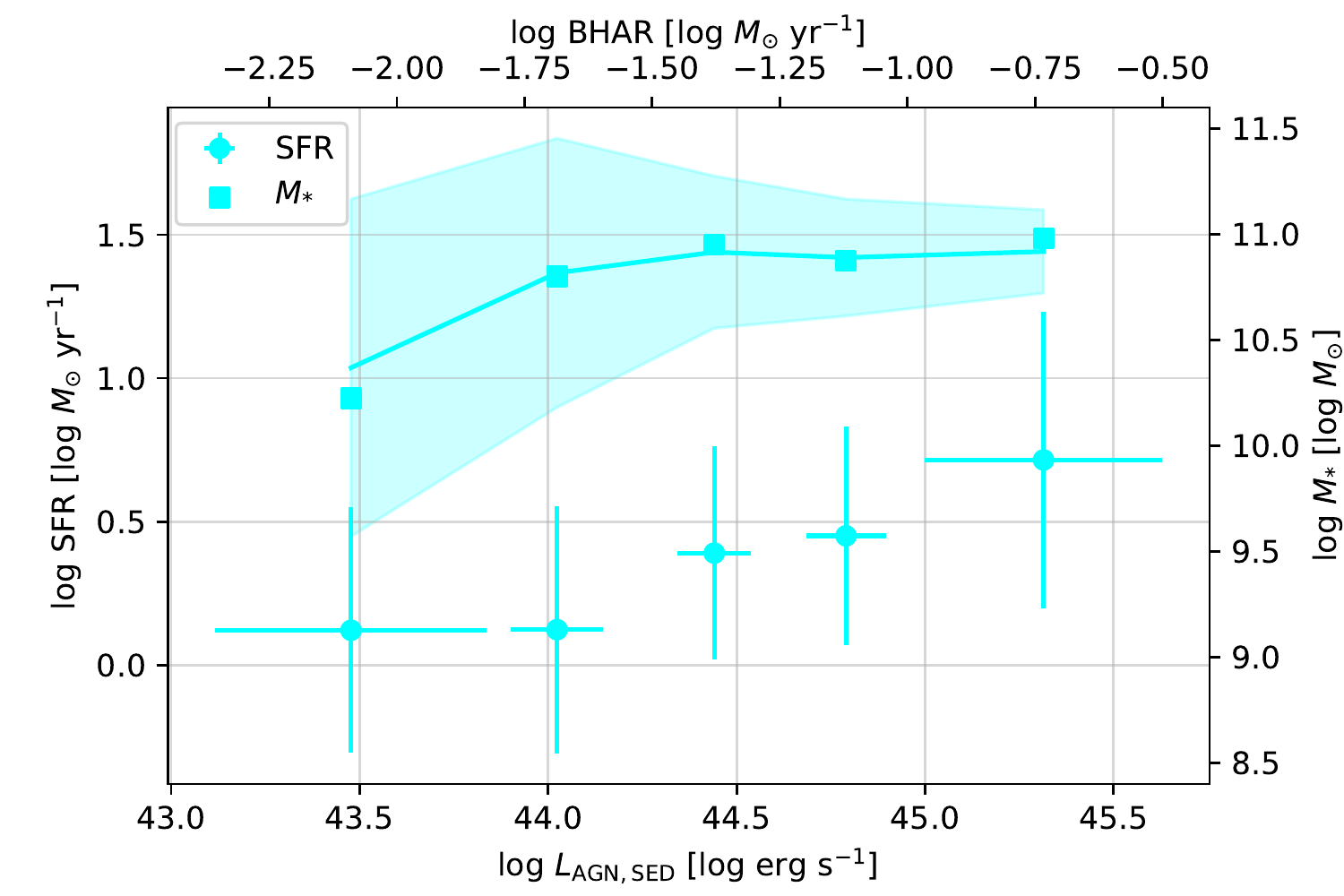}
	\includegraphics[width=0.47\textwidth]{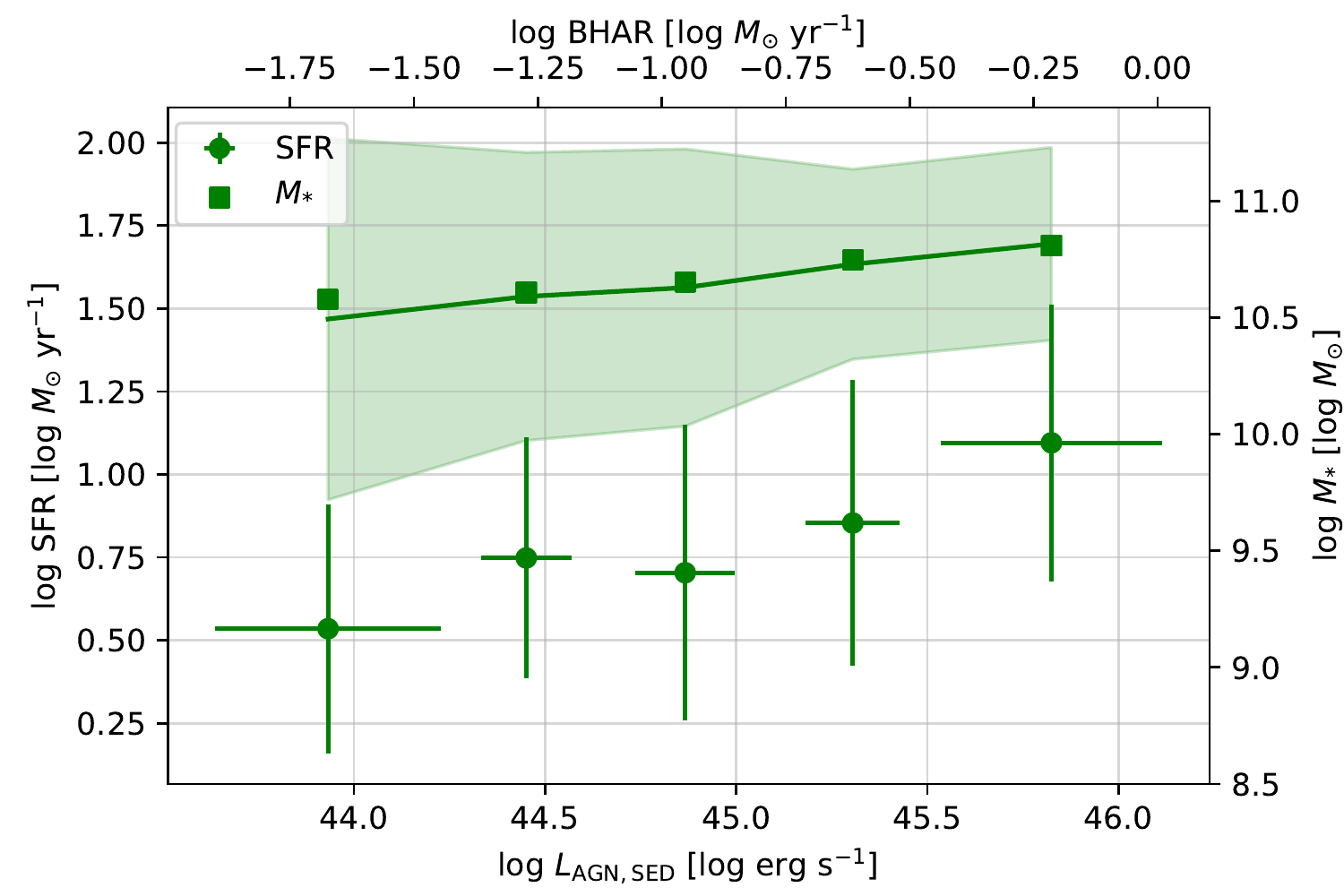}
	\includegraphics[width=0.47\textwidth]{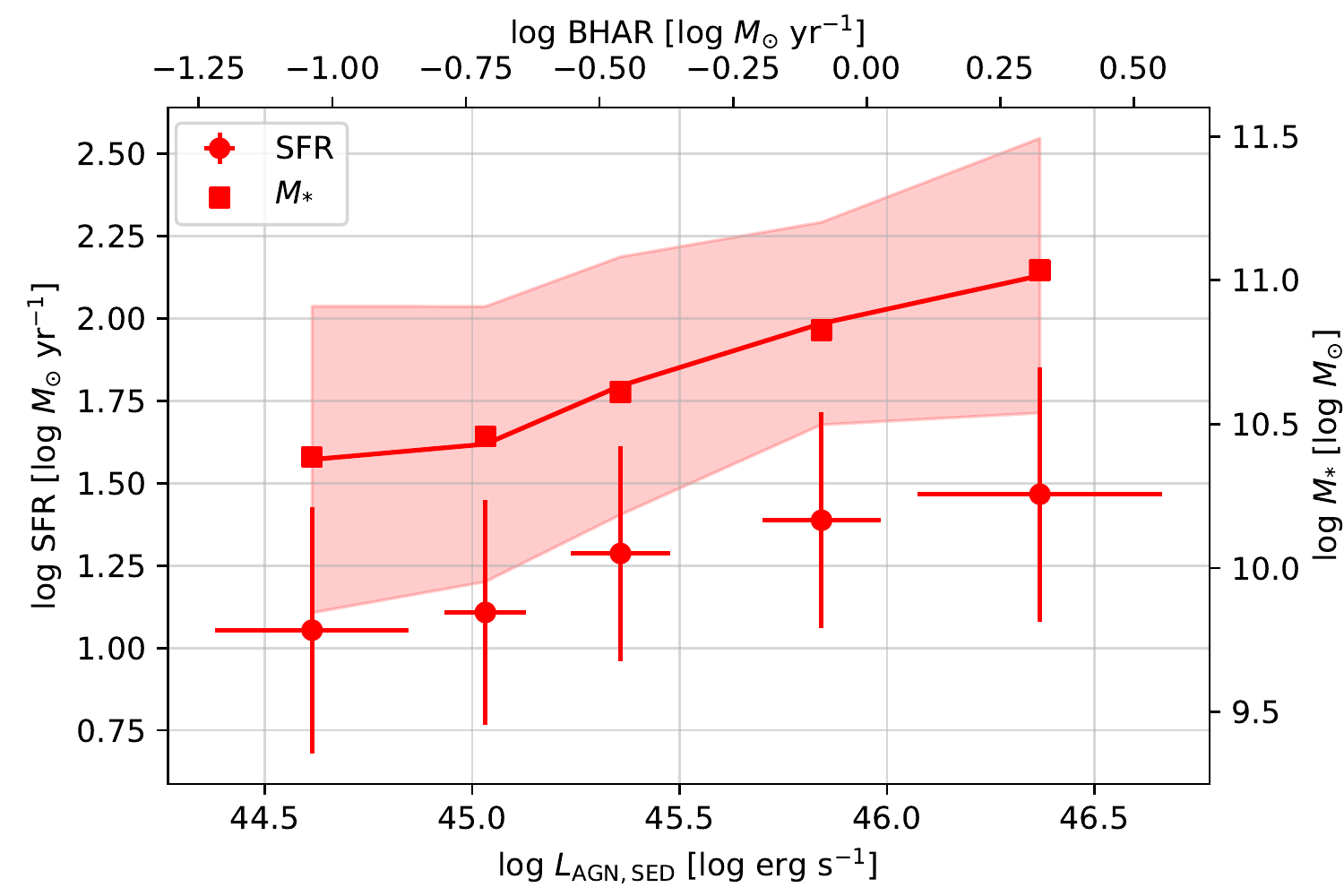} 
	
	\vspace{12pt}
	\includegraphics[width=0.94\textwidth]{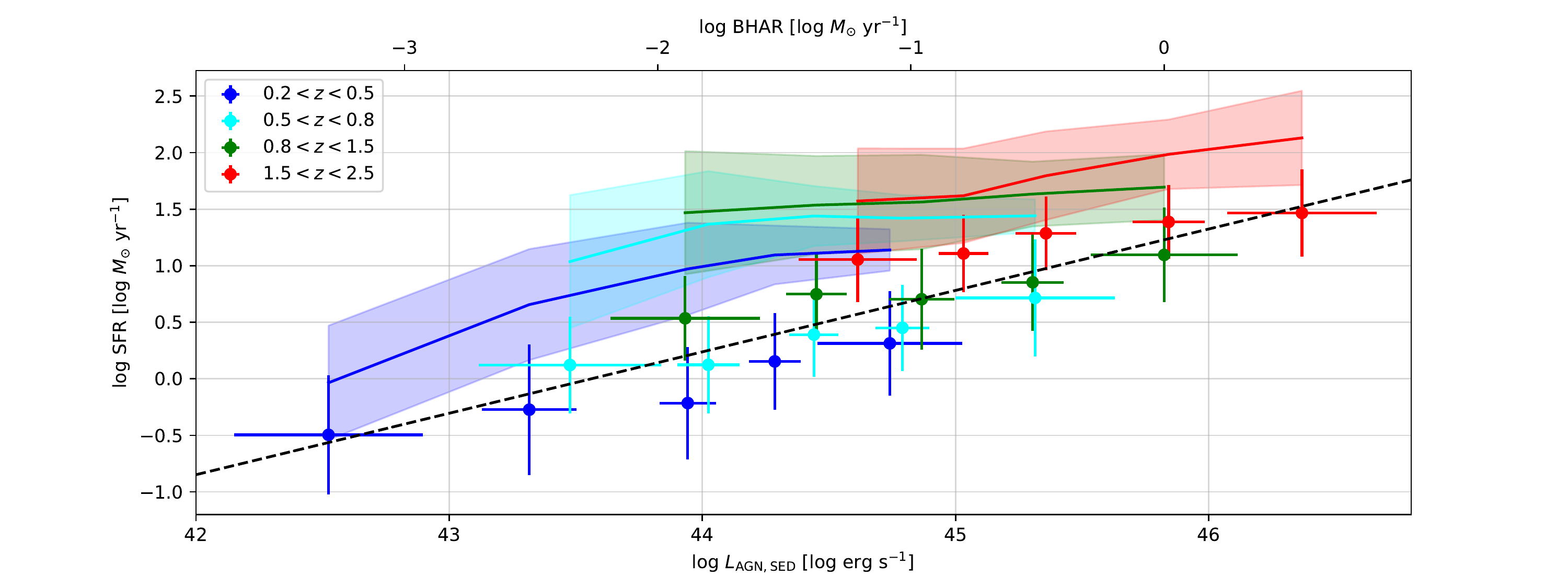}
	
	\caption{SFR as a function of AGN luminosity (BHAR) for our active galaxy sample, paneled by redshift bin and further binned by AGN luminosity; the circular markers are the data with associated errors while the solid line with shading is the calculated star-forming main sequence values (line), with SFR values calculated using the \cite{Schreiber2015} relation for star-forming main sequence galaxies with similar redshifts and stellar masses to those of our sample, and 1$\sigma$ error (shading) for each data point, while the square markers are the median stellar mass values: top left is $0.2<z<0.5$, top right is $0.5<z<0.8$, middle left is $0.8<z<1.5$, middle right is $1.5<z<2.5$; note the similar slopes between the observed and main sequence SFRs as well as the significant negative offset of the active galaxy SFRs from the main sequence. Also note the tight correlation between the median stellar mass in each bin and the main sequence SFRs. The bottom panel shows the active galaxy SFRs and main sequences for all redshift bins as a function of AGN luminosity (BHAR) as well as the line of best fit (dashed) for our data; note that redshift evolution of the relevant properties results in a steeper slope than any individual bin.}
	\label{plot: results2}
	\vspace{12pt}
\end{figure*}

\subsection{SFR and SMBH Growth are Correlated}\label{subsec: SFR Lbol link}
Previous studies examining an SFR -- AGN luminosity relation have found conflicting results. Many observational works find that SFR is correlated to AGN luminosity \citep[e.g.,][]{ Mullaney2012,Chen2013,Hickox2014,Delvecchio2015,Harris2016,Lanzuisi2017}, some find that SFR is not correlated with AGN luminosity, or at least only with a shallow relation \citep[e.g.,][]{Stanley2015, Azadi2015, Stanley2017,Shimizu2017}, while some find limited correlations. Specifically, \cite{Diamond-Stanic2012} finds that a galaxy's nuclear ($r<1$ kpc) SFR is well correlated with its AGN, but that the relation disappears when using total SFR. This is similar to results from simulations done by \cite{Volonteri2015a} examining the relation between a galaxy's SFR and the luminosity of its AGN; these simulations also predict that a galaxy's global SFR and the growth of its AGN are most strongly correlated during mergers. 

In addition, observational work by \cite{Rosario2012} finds that the dependence of a galaxy's SFR with its AGN luminosity varies, with a steep relation only existing for moderate to high luminosity AGNs ($L_{\rm AGN}>10^{44}$ erg s$^{-1}$) and low redshifts ($z<1$), with the relation flattening outside of those regimes; these findings are similar to theoretical work by \cite{Gutcke2015}. \cite{Gutcke2015} created semi-analytic models that predict that an SFR -- AGN luminosity relation evolves with AGN luminosity. Specifically, they find that the relation is flat, or even slightly negative, at low AGN luminosities and becomes steeper at higher AGN luminosities.

While some of the previous work has attempted to account for the co-dependence of SFR and AGN luminosity with redshift and stellar mass \citep[e.g.,][]{Rosario2012,Stanley2015,Azadi2015,Stanley2017}, none have done so while creating a common SED from which all AGN and galaxy parameters are derived, as this work does.

The goal of examining an SFR -- AGN luminosity relation is to understand the connection between SMBH growth and stellar mass growth in a galaxy. A more intuitive parameter for this is black hole accretion rate (BHAR) rather than AGN bolometric luminosity. We calculate BHAR for an AGN from its bolometric luminosity using the mass-energy conversion equation from \cite{Alexander2012}: BHAR $=\epsilon L_{\rm AGN}*1.5\times10^{-45}$ [($M_{\odot}$ yr$^{-1}$)/(erg s$^{-1}$)], where $\epsilon$ is the mass-energy conversion efficiency, assigned a value of 10\% ($\epsilon=0.1$) \citep{Marconi2004}. Since this is a linear scaling relation, any powerlaw slopes are equivalent between relations of SFR -- AGN luminosity and SFR -- BHAR.

In order to account for redshift evolution of the relevant galaxy properties we binned our sample by redshift, choosing similar bins to those of previous works \citep[e.g.,][]{Rosario2012,Stanley2015,Stanley2017}: $0.2<z<0.5$, $0.5<z<0.8$, $0.8<z<1.5$, and $1.5<z<2.5$. We further binned our sample by AGN bolometric luminosity in order to minimize the effect of our sample's flux-limited selection bias. We find that the SFR of an AGN host galaxy is significantly positively correlated with its AGN's bolometric luminosity, in all redshift bins and across all bins, i.e. combining all bins (see Figure \ref{plot: results2}). 

Specifically we find powerlaw slopes relating SFR to AGN bolometric luminosity (and equivalently, BHAR) of 0.36 $\pm$ 0.07, 0.34 $\pm$ 0.06, 0.27 $\pm$ 0.05, and 0.25 $\pm$ 0.03 for the redshift bins, in ascending order. These slope values are in rough agreement with the slopes of the comparative star-forming main sequence values (the shaded lines in Figure \ref{plot: results2}), with all measured SFR slopes being within 3$\sigma$ of the star-forming main sequence slope. Further, we find that the comparative star-forming main sequence tightly follows the increasing median stellar mass in each AGN bolometric luminosity bin (square datapoints in Figure \ref{plot: results2}), as expected from the \cite{Schreiber2015} relation. This indicates that an active galaxy's star formation is correlated to AGN bolometric luminosity, but may be due to a mutual dependence on galaxy stellar mass.

We do not find a slope dependence on AGN luminosity such as that found by \cite{Rosario2012} and predicted by \cite{Gutcke2015}. Instead we find that the relations are approximately linear throughout the luminosity range of our sample. Further, while these slopes show less than a one-to-one relation between SFR and BHAR, they are still significantly non-zero and are more steep than the relations seen in \cite{Stanley2015}, \cite{Azadi2015}, and \cite{Shimizu2017}. 

\begin{figure}[!t]
	\begin{center}
		\includegraphics[width=0.45\textwidth]{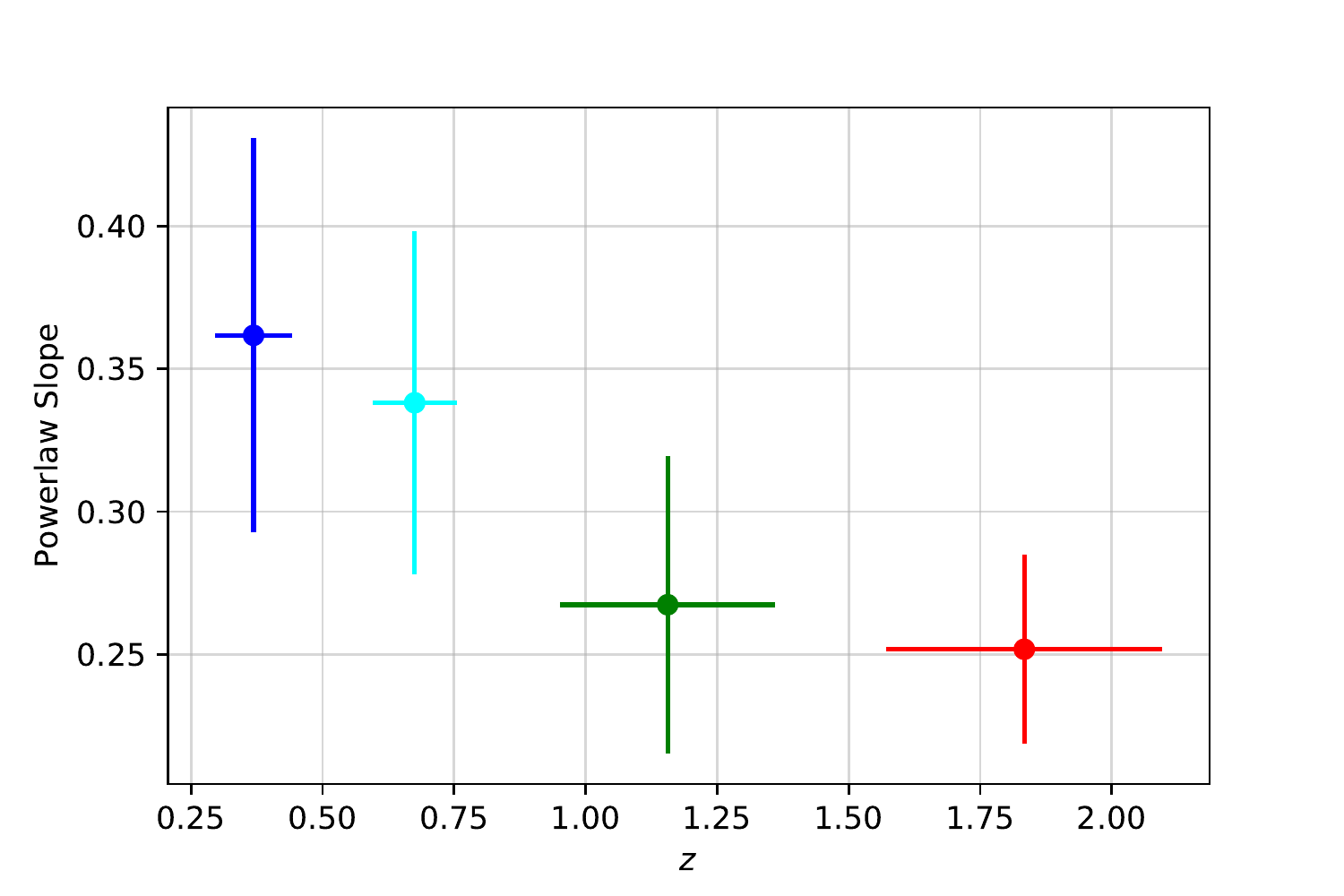}
	\end{center}
	\vspace{-12pt}
	\caption{Powerlaw slope of the SFR -- AGN luminosity relation as a function of redshift.}\label{plot: slope_evolve}
\end{figure}

We also find that the relation between SFR and AGN luminosity evolves with redshift, with the powerlaw slope of the relation flattening as redshift increases, as seen in Figure \ref{plot: slope_evolve}. The decrease in slope is gradual, with the powerlaw slope decreasing by 0.08 per unit increase in redshift. Further, with only the four redshift bins, this finding is not significant (p-value $=0.054$); however, this is an effect of underbinning our data. Examining this same evolution when using 10 equally populated redshift bins finds a similar decrease at a significant level (p-value $=0.007$). This gradual flattening of powerlaw slope with redshift is in contrast to the trend found by \cite{Stanley2017} of slope increasing as redshift increased, but agrees with findings by others \citep[e.g.,][]{Rosario2012, Azadi2015} that the relation is strongest for low redshift galaxies and flattens at higher redshifts. \cite{Rosario2012} attributes this behavior to the greater importance of major mergers, which trigger both AGNs and star formation, for galaxy and AGN growth at lower redshifts.

If we examine the relation across all the redshift bins, we find a significant powerlaw relation between SFR and AGN luminosity with a slope of $0.54\pm0.06$ and intercept of $1.23\pm0.10$. This slope is shallower than to that of \cite{Netzer2009}, which found a powerlaw relation with a slope of $\sim 0.8$ for AGN-dominated systems. This disagreement could be caused by differences in the sample arising from the observational wavelength used to select AGNs and measure SFR as was discussed in Section \ref{subsec: AGN below SFR MS}, or could arise from differences in measurements of both SFR and AGN luminosity. The work by \cite{Netzer2009} uses Oxygen emission line measurements ([OIII] and [OI]) to acquire AGN luminosity and an IR indicator, $\nu L_{\nu}$ at 60 $\mu$m, to determine SFR, while this work derives both from the same self-consistent SED model, ensuring no contamination between the SFR and AGN components. 

Our findings that SFR is positively correlated with BHAR in AGN host galaxies and that AGN host galaxies are, on average, offset below the star-forming main sequence seem to indicate that the process of AGN growth is linked to the process of global star formation in AGN host galaxies. AGN feedback is one such mechanism that could correlate these two processes.

\section{Conclusions}\label{sec: conclusions}
We present a systematic method of deriving AGN and galaxy properties by fitting multiple components to the SEDs of galaxies in multi-wavelength surveys. Using this approach on galaxies found in the ACS-GC, we create a catalog of 2585 AGN host galaxies and their properties. We analyze the AGN and host galaxy properties of this sample, with findings summarized below.

\begin{enumerate}
	\item We find that AGN host galaxies lie significantly below the star-forming main sequence, with lower SFRs than star-forming galaxies of similar mass. This offset shrinks as redshift increases. Further, we find that X-ray-selected AGN host galaxies have greater offsets from the star-forming main sequence than IR-selected AGN host galaxies. This could resolve discrepancies between previous studies about the location of AGN host galaxies relative to the star-forming main sequence, since each study selected its AGNs using different observations.
	\item We find that the SFRs of AGN host galaxies increase with AGN luminosity (and therefore BHAR) when binned by redshift, but that this may be due to a mutual dependence on galaxy stellar mass. We also find that the slope of this relation flattens as redshift increases.
\end{enumerate}

These findings suggest that a galaxy's SMBH and stellar population co-evolve, but that a process, such as AGN feedback, may restrict the SFRs of AGN host galaxies from reaching that of star-forming main sequence galaxies on average. In order to determine whether this is the case, follow-up observations are needed that identify AGN outflows and determine whether they heat or remove cool molecular gas from the host galaxy.  These are both feedback effects that could inhibit global star formation in the host galaxy.

\section*{Acknowledgments}
Support for this work was provided by NASA's Astrophysics Data Analysis program, grant number NNX15AI69G, the CU Boulder / JPL Strategic University Research Partnership program, and the National Science Foundation's Graduate Research Fellowship program. The work of DS was carried out at the Jet Propulsion Laboratory, California Institute of Technology, under a contract with NASA. RJA was supported by FONDECYT grant number 1191124. This work utilized the RMACC Summit supercomputer, which is supported by the National Science Foundation (awards ACI-1532235 and ACI-1532236), the University of Colorado Boulder, and Colorado State University. The Summit supercomputer is a joint effort of the University of Colorado Boulder and Colorado State University. The authors would also like to thank Murray Brightman, Daniel Masters, and Rebecca Nevin for providing data and support to this work.

\bibliographystyle{aasjournal}
\bibliography{library}
\end{document}